\documentclass[aps,prd,twocolumn,superscriptaddress,groupedaddress]{revtex4}
\usepackage[T1]{fontenc}
\usepackage[english]{babel}
\usepackage{amssymb}
\usepackage{cancel}
\usepackage{color,graphicx}
\usepackage{amsmath}
\usepackage{amsbsy}
\usepackage{amsthm}
\usepackage{bbm}
\usepackage{booktabs}
\usepackage{epsfig}
\usepackage{float}
\usepackage{graphicx}
\usepackage{dcolumn}
\usepackage{bbm}
\usepackage{color,epstopdf}
\usepackage{amscd}
\usepackage{amsfonts}  
\usepackage[]{amsmath}
\usepackage{amssymb}    
\usepackage{mathrsfs}
\usepackage{verbatim}
\usepackage[]{cases}
\usepackage{amsmath}
\usepackage{wasysym}
\usepackage[utf8]{inputenc}
\usepackage{mathtools}

\usepackage{bbold}
\usepackage{braket}
\usepackage{graphicx}
\begin{document}
\title{Gravitational decoherence: a non relativistic spin 1/2 fermionic model}

\author{L. Asprea}
\email{lorenzo.asprea@phd.units.it}
\affiliation{Department of Physics, University of Trieste, Strada costiera 11, 34151 Trieste, Italy}  
\affiliation{Istituto Nazionale di Fisica Nucleare, Trieste Section, Via Valerio 2, 34127 Trieste, Italy}      
\date{\today}
\author{G. Gasbarri}
\email{g.gasbarri@uab.ca}
\affiliation{F\'isica Te\`orica: Informaci\'o i Fen\`omens Qu\`antics, Department de F\'isica, Universitat Aut\`onoma de Barcelona, 08193 Bellaterra (Barcelona), Spain}
\affiliation{Department  of  Physics  and  Astronomy,University  of  Southampton,Highfield  Campus,  SO17  1BJ,United Kingdom}
\begin{abstract}
In a previous work~\cite{bosons} we have derived a quantum master equation for the dynamics of a scalar bosonic particle interacting with a weak, stochastic and classical gravitational field.  As standard matter is made of fermions, such an equation should be suitably extended to describe more relevant experimental situations. Here we derive a non relativistic model for the gravitational decoherence of spin 1/2 particles. We enrich the treatment by considering also a coupling with an external classical electromagnetic field. We comment on the differences with the scalar bosonic model and we describe the regimes in which they become negligible.
 \end{abstract}

\maketitle
\section{Introduction}
The recent exciting first detections of gravitational waves~\cite{ligo,neutronstar}, which marked a new era in astrophysics and cosmology, have pushed the scientific community towards the construction of ever more sophisticated ground and space based detectors~\cite{tian,decigo,et,kagra,lisa} to observe waves in a variety of ranges in order to sketch a first map of the stochastic gravitational background. Within the framework of quantum theory, a stochastic gravitational background affects the dynamics of matter propagation~\cite{aha,linet} and, when the quantum state is in a superposition, it leads to decoherence effects, as typical of noisy environments. Different models for the description of this phenomenon have been proposed~\cite{goku,breuer,sanchez,power,blencowe,ana,lamine}. However, they do not agree on the decoherence mechanism (the preferred basis and  rates) at which it takes place. In order to solve such apparent contradictory results, we have derived in \cite{bosons} a novel model for the decoherence effect induced by a stochastic gravitational perturbation on non relativistic scalar bosonic particles. Our model has so far proven to be able to describe more general scenarios than those present in the literature, as it is able to qualitative recover them as appropriate limiting cases, thus solving the decoherence basis puzzle. However, it might not be general enough to describe the outcome of a real experiment. The particles commonly employed in experiments (atoms, neutrons, electrons...) in fact have a charge, a spin and could be coupled to other external fields, like the Maxwell one for instance. For the above reasons, in this paper we will derive an analogous model, this time for spin 1/2 fermions interacting with both a gravitational perturbation and an external electromagnetic field.\\
\\
The paper is organized as follows. In section \ref{ii} we derive the equations of motion in Hamiltonian form for a spin 1/2 fermionic field minimally coupled to a weakly perturbed flat metric. We then specialize such equation to the non relativistic regime in section \ref{iii} and proceed with the canonical quantization of the bosonic field in the single particle sector, obtaining a Schr\"{o}dinger like equation for a test particle interacting with a weakly perturbed gravitational field.\\
In section \ref{iv} we compare the here derived fermionic model with the bosonic one derived in \cite{bosons}.\\
In section \ref{v} we specialize to the case of a stochastic gravitational perturbation and derive the corresponding master equation. We discuss the decoherence effect with explicit reference to the preferred eigenbasis and characteristic decoherence time. We also show under which assumptions our master equation is able to reproduce decoehrence in the position or momenutm eigenbasis only thus recovering the results of the literature \cite{bosons,ana,blencowe,breuer,sanchez,power}.
\section{Equations of motion}\label{ii}
We first derive the equations of motion (EOM) for a spin 1/2 fermionic field minimally coupled to linearized gravity. We start from the action for the Dirac field in curved spacetime~\cite{bd}:
\begin{equation}\label{actionf}
\begin{split}
S = \int d^4x \sqrt{-g}\mathcal{L}
\end{split}
\end{equation}
with the Lagrangian density:
\begin{equation}
\mathcal{L}  = \frac{i\hbar c}{2}[ \bar{\psi}\gamma^{\mu}{e^{\small{A}}}_\mu\mathcal{D}_{\small{A}}\psi -{e^{\small{A}}}_\mu\mathcal{D}_{\small{A}}\bar{\psi}\gamma^{\mu}\psi  ] - mc^2 \bar{\psi}\psi
\end{equation}
where ${e^A}_\mu(x)$ is the so called vierbein field \cite{vierbein}, an auxiliary field used in order to extend the definition of fermions as irreducible spin 1/2 representations of the Poincar\'{e} group to curved spacetimes (see appendix A), and 
\begin{equation}
\mathcal{D}_{\mu} \psi= \partial_{\mu}\psi +\frac{1}{8}[\gamma_a,\gamma_b]{\omega_\mu}^{ab} \psi + \frac{ie}{\hbar c}A_{\mu}\psi
\end{equation}
is the covariant derivative with respect to both the spin ($\omega_{\mu}^{ab}$) and the electromagnetic ($A_\mu$) connections.
In this framework the metric tensor $g_{AB}$ and the affine connection $\Gamma_{A}^{BC}$ are describeb by the following pair of equations:
\begin{eqnarray}\label{veir}
\left\lbrace
\begin{array}{cc}
{e_{A}}^\mu\eta_{\mu\nu}{e_{B}}^{\nu} =  g_{AB}\\
{\omega_{A}}^{\mu\nu} = {e_B}^{\mu}\eta^{\nu\rho}\partial_A {e^{B}}_\rho
+ {e_{B}}^{\mu}\eta^{\nu\rho} {e^{C}}_\rho{\Gamma^{B}}_{AC}
\end{array}
\right.
\end{eqnarray}  
Note that Eq.~\eqref{veir} holds only for a torsion free, metric compatible connection \cite{vierbein}. \\
In order to describe a weak perturbation of the metric, we now write the metric as the sum of a flat background $\eta_{\mu\nu} = \text{diag}(+---)$, and a perturbation $h_{\mu\nu}$:
\begin{equation}
g_{\mu\nu} = \eta_{\mu\nu}+h_{\mu\nu},
\end{equation}
and since we are interested in studying the dynamics of the Dirac field interacting with a weak gravitational perturbation, perform a Taylor expansion of the fermioni action around the flat background metric and truncate the series at the first perturbative order (See Appendix B for the explicit calculation). Thus, we obtain the effective Lagrangian $\mathcal{L}_{eff}$ acting on flat spacetime: 
\begin{equation}\label{effectiveaction}
\begin{split}
S &= \int d^4 x \:\Big(\frac{i\hbar c}{2}[\bar{\psi}\gamma^{\mu}\nabla_\mu\psi - \nabla_\mu(\bar{\psi})\gamma^{\mu}\psi](1+\frac{tr(h)}{2})\\
-&(1+\frac{tr(h)}{2})mc^2\bar{\psi}\psi- \frac{i\hbar c}{4}h_{\mu\nu}[\bar{\psi}\gamma^{\mu}\nabla^{\nu}\psi - \nabla^\nu(\bar{\psi})\gamma^{\mu}\psi ]\Big)\\
+& O(h^2)\\
&\equiv \int d^4 x\: \mathcal{L}_{eff} + O(h^2)
\end{split}
\end{equation}
where $\nabla_{\alpha}$ is the flat covariant derivative with respect to the electromagnetic connection. The EOM for the matter field are obtained (at first order in the perturbation $h_{\mu\nu}$) from the Euler Lagrange equations:
\begin{equation}
\frac{\partial \mathcal{L}_{eff}}{\partial \bar{\psi}} - \nabla_\alpha \frac{\partial \mathcal{L}_{eff}}{\partial \nabla_\alpha \bar{\psi} } = 0
\end{equation}
and in the harmonic gauge they read:
\begin{equation}\label{dirac}
\begin{split}
&i\hbar\partial_t\psi = eA_0\psi +  mc^2(1+\frac{h_{00}}{2})\gamma^0\psi - \frac{mc^2}{2}h_{0j}\gamma^j\psi\\
&-i\hbar c (1+\frac{h_{00}}{2})\gamma^0\gamma^i(\partial_i+\frac{ie}{\hbar c}A_i)\psi + \frac{i\hbar c}{2}h_{0i}(\partial^i+\frac{ie}{\hbar c}A^i)\psi\\
&+ \frac{i\hbar c}{2}h_{ij}\gamma^0\gamma^i(\partial^j+\frac{ie}{\hbar c}A^j)\psi+ \frac{i\hbar c}{2}h_{0i}\gamma^i\gamma^j(\partial_j+\frac{ie}{\hbar c}A_j)\psi\\
&- \frac{i\hbar c}{8}\partial_\alpha(tr(h))\gamma^0\gamma^\alpha\psi + O(h^2)\psi\\
&\quad\quad\: = : \mathcal{H}\psi +O(h^2)\psi
\end{split}
\end{equation}
As in the case of a scalar field  studied in~\cite{bosons} , we are not allowed to give a probabilistic interpretation to the field $\psi$, becuse the conserved charged $Q$ associated to the internal $U(1)$ symmetry ($\psi \rightarrow e^{ie}\psi$ ; $\bar{\psi}\rightarrow e^{-ie}\bar{\psi}$) via Noether's Theorem reads:
\begin{equation}\label{chargef}
\begin{split}
Q &\equiv -ie \int d^3 x \: \Big(\frac{\partial\mathcal{L}_{eff}}{\partial(\nabla_0 \psi)}\psi - \psi^\dagger \frac{\partial\mathcal{L}_{eff}}{\partial(\nabla_0 \psi^\dagger)}\Big)\\
&= \hbar e c \int d^3 x\: \Big( \psi^\dagger (1-tr(h)-\frac{h_{00}}{2})\psi - \psi^\dagger \frac{h_{0i}}{2}\gamma^0\gamma^i\psi  \Big)
\end{split}
\end{equation}
instead of the required:
\begin{equation}\label{eq:stchf}
\rho = \int d^3 x \:
\psi^\dagger \psi
\end{equation}
We therefore apply the transformation:
\begin{eqnarray}\label{eq:chargpres}
\left\lbrace
\begin{array}{ccc}
T &=& (1 -\frac{tr(h)}{2}-\frac{h_{00}}{4} -\frac{h_{0i}}{4}\gamma^0\gamma^i )\\
 \psi &\rightarrow& T \psi \\
\mathcal{H}&\rightarrow& \mathfrak{H} := T \mathcal{H}T^{-1}+ i\hbar T \partial_t(T^{-1})
\end{array}
\right.
\end{eqnarray}
so that, in the new representation,the conserved charge can be expressed by the standard form in Eq.~\eqref{eq:stchf}. \\
After some algebra the EOM~\eqref{dirac} read:
\begin{equation}\label{eqmotf}
i\hbar\partial_t \psi = [mc^2\gamma^0 + \mathfrak{E}+ \mathcal{O}]\psi
\end{equation}
where
\begin{equation}
\begin{split}
\mathfrak{E} = & e A_0 +\frac{mc^2}{2}h_{00}\gamma^0 + i\hbar c \: h_{0i} (\partial^i-\frac{ie}{\hbar c}A^i)+\frac{i\hbar c}{4}\partial_{i}(h_0^i)\\
&+\frac{\hbar c}{4}\epsilon^{ijk}\partial_i(h_{0j})\Sigma_k-\frac{3i\hbar}{8}\partial_t(tr(h))+\frac{i\hbar}{4}\partial_t(h_{00})
\end{split}
\end{equation}
\begin{equation}
\begin{split}
\mathcal{O} =& - i\hbar c (1+\frac{h_{00}}{2}) (\partial_j - \frac{ie}{\hbar c}A_j)\alpha^j +\frac{i\hbar}{4}\partial_t(h_{0i})\alpha^i \\
&+ \frac{i\hbar c}{2}h_{ij}(\partial^j - \frac{ie}{\hbar c}A^j)\alpha^i + \frac{i\hbar c}{4}\partial_i(\frac{tr(h)}{2}-h_{00})\alpha^i      
\end{split}
\end{equation}
are respectively the even (diagonal) and odd (off diagonal) parts of the Hamiltonian $\mathfrak{H}$, with $\alpha^{\mu}=\gamma^0\gamma^\mu$ and $\Sigma^i = \textrm{diag}(\sigma^i,\sigma^i)$.\\
We are interested in the description of the dynamics of a positive energy particle system in the non relativistic limit. In such a limit, the particle and antiparticle sectors are non interacting with one another, that is to say, the EOM~\eqref{dirac} can be recast to a system of two decoupled equations respectively for the large ($\psi_L$) and the small ($\psi_s$) component of the bispinor $\psi = \left(\begin{array}{cc} \psi_L \\ \psi_s \end{array}\right)$. While this is evident and straightforward for the free case \cite{greiner}, for an interacting theory decoupling the two components is a very complicated task that can only be achieved perturbatively.\\
\\
In the next section we will provide a standard prescription for the diagonalization of the EOM in the non relativistic limit.
\section{Non relativistic limit and canonical quantization}\label{iii}
We aim to find a representation of the bispinor field $\psi$ in which the EOM \eqref{eqmotf} are diagonal. This representation can be found in non relativistic limit
following the Foldy-Wouthuysen Method~\cite{foldy}, which allows one to write perturbatively (at any order in $\frac{v}{c}$) two decoupled equations, one for each component of the field. The method is operatively characterized by the application of an appropriate unitary transformation $U$:
\begin{equation}
\psi  \rightarrow  \psi^\prime = U\psi 
\end{equation}
\begin{equation}
\begin{split}
\mathfrak{H} \rightarrow  \mathfrak{H}^\prime =& \:U(\mathfrak{H}-i\hbar\partial_t)U^{-1}\\
 =& \: mc^2 \gamma^0+ \mathfrak{E}^{\prime} + \mathcal{O}^{\prime} +O(h^2)
\end{split}
\end{equation}
such that, in the new representation, the antidiagonal part $\mathcal{O}^\prime$ is of higher order in $\frac{v}{c}$ than the diagonal $\mathfrak{E}^\prime$. By neglecting $\mathcal{O}^\prime$ one recovers two decoupled equations. 
By performing iteratively the transformation, one can always find a representation of the bispinor field for which the EOM are diagonal at any desired order in $\frac{v}{c}$.\\
In our case, the task is easily achieved by applying the subsequent transformations:
\begin{eqnarray}\label{transf}
\left\{
\begin{array}{ccc}
U =& e^{- i \gamma^0 \mathcal{O} /(2mc^2)} \\
U^\prime =& e^{-i\gamma^0 \mathcal{O^\prime} /(2mc^2)}\\
U^{\prime\prime} =& e^{-i\gamma^0 \mathcal{O^{\prime\prime}} /(2mc^2)}
\end{array}
\right.
\end{eqnarray}
after which, with some algebra (see Appendix C) and by neglecting the terms containing the derivatives of the gravitational perturbation of order $\frac{v^3}{c^3}$ or higher\footnote{For the sake of compactness we relegate such terms to Appendix D. Note also that in most experimental situations such contributions are negligible in the case of gravity, and they wouldn't add any further informative content to the analysis in any case.}, the Hamiltonian density to order $\frac{v^4}{c^4}$ reads:
\begin{widetext}
\begin{equation}\label{hamf}
\begin{split}
    H=&  eA_0+\gamma^{0}\Bigg[mc^2(1+\frac{h_{00}}{2})-\frac{\hbar^2}{2m}(1+\frac{h_{00}}{2})(\mathbf{\nabla}-\frac{ie}{\hbar c}\mathbf{A})^2 -\frac{\hbar e}{2m c}(1+\frac{h_{00}}{2})B^k\Sigma_k -\frac{\hbar^2}{2m}h_{ij}(\partial^i-\frac{ie}{\hbar c}A^i)(\partial^j-\frac{ie}{\hbar c}A^j)\\
    &+\frac{\hbar e}{4 mc}\epsilon^{ijl}h_{jk}{F_i}^k\Sigma_l \Bigg] +\frac{i\hbar^2 e}{4m^2c^2}(1+\frac{h_{00}}{2})\Big( \frac{\mathbf{\nabla}}{2}\times\mathbf{E} - \mathbf{E}\times\mathbf{\nabla}\Big)\cdot\boldsymbol{\Sigma} - (1+h_{00})\frac{\hbar^2e}{8m^2c^2}\mathbf{\nabla}\cdot\mathbf{E}\\
    &-\frac{i\hbar^2 e}{16m^2c^2}\epsilon^{ikl}h_{ij}\partial^{j}(E_k)\Sigma_l -\frac{i\hbar^2e}{8m^2c^2}\epsilon^{ikl}h_{ij}E_k(\partial^j-\frac{ie}{\hbar c}A^j)\Sigma_l + \frac{i\hbar^2 e}{4m^2c^2}\epsilon^{ijl}h_{0k}{F_j}^k(\partial_i-\frac{ie}{\hbar c}A_i)\Sigma_l\\
    &-\frac{\hbar^2 e}{8m^2c^2}h_{0j}\partial_i(F^{ij})+\frac{i\hbar^2 e}{8m^2c^2}\epsilon^{ijl}h_{0k}\partial_i({F_j}^k)\Sigma_l-\frac{\gamma^0}{8m^3c^6}\Bigg[\hbar^4c^4(1+2h_{00})(\mathbf{\nabla}-\frac{ie}{\hbar c}\mathbf{A})^4 + \hbar^2 e c^2 (1+2h_{00})B^2 \\
&+ 2 \hbar^4 c^4 h_{ij}(\mathbf{\nabla}-\frac{ie}{\hbar c}\mathbf{A})^2(\partial^i-\frac{ie}{\hbar c }A^i)(\partial^j-\frac{ie}{\hbar c }A^j) - \frac{\hbar^3 e c^3}{2}\epsilon^{ijl}h_{jm}{F_i}^mB^k\lbrace \Sigma_k ,\Sigma_l\rbrace\\
& +\frac{\hbar^3 e c^3}{2}\epsilon^{ijl}\lbrace (\mathbf{\nabla}-\frac{ie}{\hbar c}\mathbf{A})^2 , h_{jk}{F_i}^k \rbrace \Sigma_l-\hbar^3 e c^3 (1+2h_{00})\lbrace(\mathbf{\nabla}-\frac{ie}{\hbar c}\mathbf{A})^2,B^k \rbrace\Sigma_k \Bigg]\\
&+H_d+O(h^2)+O(\partial h) +O(\frac{v^5}{c^5})
\end{split}
\end{equation}
\end{widetext}
where $B$ and $E$ are the magnetic and electric field, and in terms of the four-potential they read:
\begin{eqnarray}
\left\lbrace
\begin{array}{cc}
\mathbf{E} =& - \mathbf{\nabla} A_0 - \frac{1}{c}\dot{\textbf{A}}\\
\mathbf{B} =& \mathbf{\nabla}\times\mathbf{A}\\
B^k =& -\frac{1}{2} \epsilon^{ijk}F_{ij}\\
F_{ij} =& - \epsilon_{ijk}B^{k}
\end{array}
\right.
\end{eqnarray}
$\epsilon_{ijk}$ represent the Levi-Civita symbol, and
\begin{equation}
\begin{split}
H_d= &-\frac{\hbar^2}{8m}\partial_i(h_{00})(\partial^i-\frac{ie}{\hbar c}A^i)\gamma^0 -\frac{\hbar^2}{16m}\partial^i\partial_i(h_{00})\gamma^0 \\
&+\frac{i\hbar c}{4}\partial_{i}(h_0^i)+\frac{\hbar c}{4}\epsilon^{ijk}\partial_i(h_{0j})\Sigma_k -\frac{3i\hbar}{8}\partial_t(tr(h))\\
&+\frac{i\hbar}{4}\partial_t(h_{00}) +\gamma^0 \Bigg[ \frac{\hbar^2}{2m}\partial^i(h_{00})\nabla_i -\frac{\hbar^2}{4m}\partial^i(h_{ij})\nabla^j\\ &-\frac{\hbar^2}{2m}\partial_i(\frac{tr(h)}{2}-h_{00})\nabla^i-\frac{\hbar^2}{4m}\partial^i\partial_i (\frac{tr(h)}{2}-h_{00})\\
&-\frac{i\hbar^2}{4m}\epsilon^{ijk}\Big(\partial_i(h_{00})\nabla_j-\partial_i(h_{jl})\nabla^l \Big)\Sigma_k \Bigg]    
\end{split}    
\end{equation}
Note that as the transformations \eqref{transf} are unitary~\cite{armin}, they preserve the conserved charge in \eqref{chargef}, i.e. the probability density in the non relativistic limit.\\ 
In the non relativistic limit the EOM \eqref{hamf} do not mix the two components $\psi_L$ and $\psi_s$  of the field (up to a very small correction). As we are interested in the dynamics of particles only, we restrict the analysis to the first field component $\psi_L$, that we rename as $\psi$ in what follows.\\ 
Since the dynamics preserves the probability density, we are allowed to apply the canonical quantization prescription and impose the equal time commutation relations:
\begin{equation}
\begin{split}
[\hat{\psi}(t,\mathbf{x}),\hat{\psi}(t,\mathbf{x'})] =  & [\hat{\psi}^{\dagger}(t,\mathbf{x}),\hat{\psi}^{\dagger}(t,\mathbf{x'})] =0\\
[\hat{\psi}(t,\mathbf{x}),\hat{\psi}^{\dagger}(t,\mathbf{x'})] =& \: \delta^{3}(\mathbf{x}-\mathbf{x'})
\end{split}
\end{equation}
to obtain the EOM for the quantum field. The equation thus obtained does not allow for the creation or annihilation of particles. We can thus safely project it onto a single particle sector to obtain the single particle Schr\"{o}dinger like equation:
\begin{equation}\label{schrf}
i\hbar\partial_t\vert\phi(t)\rangle = (\hat{H}_0+\hat{H}_r+\hat{H}_p+\hat{H}_{rp}+\hat{H}_d)\vert\phi(t)\rangle
\end{equation}
with:
\begin{equation}
\hat{H}_{0} = mc^2 +\frac{1}{2m}\Big(\hat{\mathbf{p}}-\frac{e}{c}\mathbf{A}(\hat{\mathbf{x}})\Big)^2+eA_0(\hat{\mathbf{x}})- \frac{\hbar e}{2mc}\mathbf{B}(\hat{\mathbf{x}})\cdot\boldsymbol{\sigma}    
\end{equation}
\begin{equation}
\begin{split}
\hat{H}_r =& \frac{\hbar e}{4m^2c^2}\Big( \frac{\hat{\mathbf{p}}}{2}\times\mathbf{E}(\hat{\mathbf{x}}) - \mathbf{E}(\hat{\mathbf{x}})\times\hat{\mathbf{p}}\Big)\cdot\boldsymbol{\sigma} \\
&- \frac{\hbar^2e}{8m^2c^2}\mathbf{\nabla}\cdot\mathbf{E}(\hat{\mathbf{x}})-\frac{\gamma^0}{8m^3c^6}\Bigg[c^4(\hat{\mathbf{p}}-\frac{e}{ c}\mathbf{A}(\hat{\mathbf{x}}))^4 \\
&+ \hbar^2 e c^2 B^2(\hat{\mathbf{x}})-\hbar e c^3 \lbrace(\hat{\mathbf{p}}-\frac{e}{ c}\mathbf{A}(\hat{\mathbf{x}}))^2,B^k(\hat{\mathbf{x}}) \rbrace\sigma_k \Bigg]    
\end{split}    
\end{equation}
\begin{equation}
\begin{split}
\hat{H}_p =& +\frac{mc^2}{2}h^{00}(t,\hat{\mathbf{x}})- \frac{1}{8m}\lbrace h^{00}(t,\hat{\mathbf{x}})\:,\:\Big(\hat{\mathbf{p}}-\frac{e}{c}\mathbf{A}(\hat{\mathbf{x}})\Big)^2\rbrace \\
&+ \frac{c}{2} \lbrace h_{0i}(t,\hat{\mathbf{x}})\: ,\: \hat{p}^i\rbrace -\frac{\hbar e }{4mc}\epsilon^{ikl}h_{ij}(\hat{\mathbf{x}},t){F^j}_k(\hat{\mathbf{x}})\sigma_l \\
&-\frac{1}{4m}\lbrace h^{ij}(t,\hat{\mathbf{x}})\: ,\:\Big(\hat{p}_i-\frac{e}{c}A_i(t,\hat{\mathbf{x}})\Big)\Big(\hat{p}_j-\frac{e}{c}A_j(t,\hat{\mathbf{x}})\Big)\rbrace\\
&- \frac{\hbar e}{4mc}h_{00}(\hat{\mathbf{x}},t)\mathbf{B}(\hat{\mathbf{x}})\cdot\boldsymbol{\sigma}
\end{split}
\end{equation}
\begin{widetext}
\begin{equation}
 \begin{split}
\hat{H}_{rp}=& \frac{\hbar e}{16m^2c^2}\lbrace h_{00}(\hat{\mathbf{x}},t),\Big( \frac{\hat{\mathbf{p}}}{2}\times\mathbf{E}(\hat{\mathbf{x}}) - \mathbf{E}(\hat{\mathbf{x}})\times\hat{\mathbf{p}}\Big)\cdot\boldsymbol{\sigma}\rbrace -\frac{i\hbar^2 e}{16m^2c^2}\epsilon^{ikl} h_{ij}(\hat{\mathbf{x}})\partial^{j}(E_k(\hat{\mathbf{x}}))\Sigma_l - \frac{\hbar^2e}{8m^2c^2}h_{00}(\hat{\mathbf{x}})(\hat{\mathbf{x}},t)\mathbf{\nabla}\cdot\mathbf{E}(\hat{\mathbf{x}})\\
&-\frac{\hbar e}{16m^2c^2}\epsilon^{ikl}\lbrace h_{ij}(\hat{\mathbf{x}},t),E_k(\hat{\mathbf{x}})(\hat{p}^j-\frac{e}{ c}A^j(\hat{\mathbf{x}}))\rbrace\Sigma_l+ \frac{\hbar e}{8m^2c^2}\epsilon^{ijl}\lbrace h_{0k}(\hat{\mathbf{x}}),{F_j}^k(\hat{p}_i-\frac{e}{ c}A_i)\rbrace\Sigma_l\\
&-\frac{\hbar^2 e}{8m^2c^2}h_{0j}(\hat{\mathbf{x}})\partial_i(F^{ij}(\hat{\mathbf{x}}))+\frac{i\hbar^2 e}{8m^2c^2}\epsilon^{ijl}h_{0k}(\hat{\mathbf{x}})\partial_i({F_j}^k(\hat{\mathbf{x}}))\Sigma_l-\frac{\gamma^0}{8m^3c^6}\Bigg[c^4\lbrace h_{00}(\hat{\mathbf{x}}),(\mathbf{\nabla}-\frac{e}{ c}\mathbf{A}(\hat{\mathbf{x}}))^4 \\
& +  c^4 \lbrace h_{ij}(\hat{\mathbf{x}}),(\hat{\mathbf{p}}-\frac{e}{ c}\mathbf{A}(\hat{\mathbf{x}}))^2(\hat{p}^i-\frac{e}{ c }A^i(\hat{\mathbf{x}}))(\hat{p}^j-\frac{e}{ c }A^j(\hat{\mathbf{x}}))\rbrace - \frac{\hbar^3 e c^3}{2}\epsilon^{ijl}h_{jm}(\hat{\mathbf{x}}){F_i}^m(\hat{\mathbf{x}})B^k(\hat{\mathbf{x}})\lbrace \Sigma_k ,\Sigma_l\rbrace \\
&+ 2\hbar^2 e c^2 h_{00}(\hat{\mathbf{x}})B^2(\hat{\mathbf{x}})+\frac{\hbar e c^3}{2}\epsilon^{ijl}\lbrace (\hat{\mathbf{p}}-\frac{e}{ c}\mathbf{A})^2 , h_{jk}(\hat{\mathbf{x}}){F_i}^k(\hat{\mathbf{x}}) \rbrace \Sigma_l-\hbar e c^3 \lbrace h_{00}(\hat{\mathbf{x}})\lbrace(\hat{\mathbf{p}}-\frac{e}{c}\mathbf{A})^2,B^k \rbrace\rbrace\Sigma_k \Bigg]\\
\end{split}
\end{equation}
\begin{equation}
\begin{split}
\hat{H}_d =&  -\frac{\hbar}{16m}\lbrace\partial_i(h_{00}(\hat{\mathbf{x}})),(\hat{\mathbf{p}}^i-\frac{e}{ c}A^i(\hat{\mathbf{x}}))\rbrace\gamma^0 -\frac{\hbar^2}{16m}\partial^i\partial_i(h_{00}(\hat{\mathbf{x}}))\gamma^0 +\frac{i\hbar c}{4}\partial_{i}(h_0^i(\hat{\mathbf{x}}))+\frac{\hbar c}{4}\epsilon^{ijk}\partial_i(h_{0j}(\hat{\mathbf{x}}))\sigma_k\\ &-\frac{3i\hbar}{8}\partial_t(tr(h(\hat{\mathbf{x}})))+\frac{i\hbar}{4}\partial_t(h_{00}(\hat{\mathbf{x}})) +\gamma^0 \Bigg[ \frac{\hbar^2}{4m}\lbrace\partial^i(h_{00}(\hat{\mathbf{x}})),(\hat{\mathbf{p}}_i-\frac{e}{c}A_i(\hat{\mathbf{x}}))\rbrace -\frac{\hbar^2}{8m}\lbrace\partial^i(h_{ij}(\hat{\mathbf{x}})),(\hat{\mathbf{p}}^j-\frac{e}{c}A^j(\hat{\mathbf{x}}))\rbrace \\ &-\frac{\hbar^2}{4m}\lbrace\partial_i(\frac{tr(h(\hat{\mathbf{x}}))}{2}-h_{00}(\hat{\mathbf{x}})),(\hat{\mathbf{p}}^i-\frac{e}{c}A^i(\hat{\mathbf{x}}))\rbrace-\frac{\hbar^2}{4m}\partial^i\partial_i (\frac{tr(h(\hat{\mathbf{x}}))}{2}-h_{00}(\hat{\mathbf{x}}))-\frac{e}{c}A^l(\hat{\mathbf{x}}))\rbrace \Big)\sigma_k\\
&-\frac{i\hbar^2}{8m}\epsilon^{ijk}\Big(\lbrace\partial_i(h_{00}(\hat{\mathbf{x}})),(\hat{\mathbf{p}}_j-\frac{e}{c}A_j(\hat{\mathbf{x}}))\rbrace-\lbrace\partial_i(h_{jl}(\hat{\mathbf{x}})),\hat{\mathbf{p}}^l-\frac{e}{c}A^l(\hat{\mathbf{x}})\rbrace \Bigg]   
\end{split}
\end{equation}
\end{widetext}
where $\hat{\mathbf{x}},\hat{\mathbf{p}}$ are respectively the single particle  position and the momentum operator. The term $\hat{H}_0$ is the usual Pauli Hamiltonian \cite{greiner} plus an irrelevant global phase $mc^2$ that can be reabsorbed with the transformation:
\begin{equation}
\vert\phi (t)\rangle \rightarrow e^{imc^2 t/\hbar}\vert\phi (t)\rangle
\end{equation} 
The term $\hat{H}_r$ encodes the standard relativistic corrections~\cite{armin} due to the presence of the electromagnetic field up to order $\frac{v^4}{c^4}$. Finally, $\hat{H}_p$, $\hat{H}_d$ and $\hat{H}_{rp}$ account for the corrections due to the presence of the weak gravitational field respectively to $\hat{H}_0$ and $\hat{H}_r$.\\
\section{Differences with the bosonic model}\label{iv}
Equation~\eqref{hamf} is rather instructive as it extends the non relativistic Hamiltonian obtained in \cite{bosons} for the scalar field (namely Eq.(21) of \cite{bosons}) which we recall below~\footnote{We have added the superscripts (B) and (F) for respectively bosonic and fermionic in order to avoid confusion}:
\begin{equation}\label{hamB}
\begin{split}
{\hat{H}_0}^{(B)} =& mc^2 +\frac{\hat{\mathbf{p}}^2}{2m}\\
{\hat{H}_p}^{(B)} =& \frac{mc^2}{2}h^{00}(t,\hat{\mathbf{x}}) - \frac{\hbar^2}{8m}\lbrace h^{00}(t,\hat{\mathbf{x}}),\hat{\mathbf{p}}^2\rbrace + \frac{c}{2}\lbrace h^{0i}(t,\hat{\mathbf{x}}),\hat{p}_i\rbrace\\ &-\frac{1}{4m}\lbrace h^{ij}(t,\hat{\mathbf{x}}),\hat{p}_i\hat{p}_j\rbrace \\
{\hat{H}_d}^{(B)}=&\frac{\hbar^2}{8m}\mathbf{\nabla}^2(tr[h^{\mu\nu}(t,\hat{\mathbf{x}})])+\frac{i\hbar}{2}\partial_t(h^{00}(t,\hat{\mathbf{x}}))\\
&-\frac{i\hbar}{4}\partial_t(tr[h^{\mu\nu}(t,\hat{\mathbf{x}})])
\end{split}
\end{equation}
to describe the dynamics of a charged quantum particle (with spin 1/2) subject to a gravitational perturbation and to an external electromagnetic field. It is however as instructive to consider the electromagnetic free case , i.e. the limit $A(t,\hat{\mathbf{x}})\rightarrow 0$, in order to directly compare the fermionic and bosonic Hamiltonian. By taking the limit $A(t,\hat{\mathbf{x}})\rightarrow 0$ of Eq.~\eqref{hamf}, we obtain:
\begin{equation}\label{hamilF}
\begin{split}
{\hat{H}_0}^{(F)} =& mc^2 +\frac{\hat{\mathbf{p}}^2}{2m}\\
{\hat{H}_p}^{(F)} =& \frac{mc^2}{2}h^{00}(t,\hat{\mathbf{x}}) - \frac{\hbar^2}{8m}\lbrace h^{00}(t,\hat{\mathbf{x}}),\hat{\mathbf{p}}^2\rbrace + \frac{c}{2}\lbrace h^{0i}(t,\hat{\mathbf{x}}),\hat{p}_i\rbrace\\ &-\frac{1}{4m}\lbrace h^{ij}(t,\hat{\mathbf{x}}),\hat{p}_i\hat{p}_j\rbrace 
\end{split}
\end{equation}
\begin{equation}
\begin{split}
&{\hat{H}_d}^{(F)}= -\frac{\hbar}{16m}\lbrace\partial_i(h_{00}(\hat{\mathbf{x}})),(\hat{\mathbf{p}}^i-\frac{e}{ c}A^i(\hat{\mathbf{x}}))\rbrace\gamma^0\\ &-\frac{\hbar^2}{16m}\partial^i\partial_i(h_{00}(\hat{\mathbf{x}}))\gamma^0 +\frac{i\hbar c}{4}\partial_{i}(h_0^i(\hat{\mathbf{x}}))-\frac{3i\hbar}{8}\partial_t(tr(h(\hat{\mathbf{x}})))\\
&+\frac{\hbar c}{4}\epsilon^{ijk}\partial_i(h_{0j}(\hat{\mathbf{x}}))\sigma_k +\frac{i\hbar}{4}\partial_t(h_{00}(\hat{\mathbf{x}}))\\
&+\gamma^0 \Bigg[ \frac{\hbar^2}{4m}\lbrace\partial^i(h_{00}(\hat{\mathbf{x}})),(\hat{\mathbf{p}}_i-\frac{e}{c}A_i(\hat{\mathbf{x}}))\rbrace \\
&-\frac{\hbar^2}{8m}\lbrace\partial^i(h_{ij}(\hat{\mathbf{x}})),(\hat{\mathbf{p}}^j-\frac{e}{c}A^j(\hat{\mathbf{x}}))\rbrace \\ &-\frac{\hbar^2}{4m}\lbrace\partial_i(\frac{tr(h(\hat{\mathbf{x}}))}{2}-h_{00}(\hat{\mathbf{x}})),(\hat{\mathbf{p}}^i-\frac{e}{c}A^i(\hat{\mathbf{x}}))\rbrace\\
&-\frac{\hbar^2}{4m}\partial^i\partial_i (\frac{tr(h(\hat{\mathbf{x}}))}{2}-h_{00}(\hat{\mathbf{x}}))\\
&-\frac{i\hbar^2}{8m}\epsilon^{ijk}\Big(\lbrace\partial_i(h_{00}(\hat{\mathbf{x}})),(\hat{\mathbf{p}}_j-\frac{e}{c}A_j(\hat{\mathbf{x}}))\rbrace\\
&-\lbrace\partial_i(h_{jl}(\hat{\mathbf{x}})),(\hat{\mathbf{p}}^l-\frac{e}{c}A^l(\hat{\mathbf{x}}))\rbrace \Big)\sigma_k \Bigg]
\end{split}
\end{equation}
As expected, the bosonic and the fermionc description match for the gravity free case (${\hat{H}_0}^{(B)}={\hat{H}_0}^{(F)}$). They also match for the terms proportional to the gravitational perturbation $h_{\mu\nu}$. This is also to be expected: suppose in fact that there were actually a difference in the terms $h_{00}\hat{\mathbf{p}}^2/2m$ or $mc^2h_{00}$. This would imply that e.g. a simple change form Cartesian to Rindler \cite{mtw} coordinates would predict for a boson and a fermion to fall with the same acceleration in the first (Cartesian) but not the second (Rindler) reference frame, which would violate the weak equivalence principle.
The same line of reasoning can be applied to the other terms containing $h_{ij}$and $h_{0i}$. It is interesting however to notice that some differences arise for the terms containing the derivatives of the gravitational perturbation $\partial h_{\mu\nu}$. Such differences originated when we required the matter field to allow for a probabilistic interpretation in order to canonically quantize the system (see Eq. (\ref{eq:chargpres}) and Eq. (x) of \cite{bosons}).  
\section{Master equation with electromagnetic field}\label{v}
In this section we derive a master equation to describe the decoherence effect induced by a weak stochastic gravitational perturbation on a spin 1/2 fermionic particle, as done for the scalar case in the previous chapter. For the sake of simplicity and compactness of the result, we will restrict our analysis to the Pauli Hamiltonian $\hat{H}_0$ and its gravitational corrections $\hat{H}_p$, as the terms $\hat{H}_r$ and $\hat{H}_{rp}$ are of higher order in the non relativistic expansion \footnote{Note however that one needs to be careful when applying the results of this section to a real experimental situations, as the term $\hat{H}_r$ might dominate over $\hat{H}_p$ (depending on the size of $\mathbf{E}$,$\mathbf{B}$, and $h_{\mu\nu}$) and should therefore be taken in consideration.}, and the term $\hat{H}_d$ contains derivatives of the gravitational perturbations, as in typical experimental situations~\cite{ligo,tian,decigo,et,kagra} they are negligible and would not add any further informative content to the analysis in any case.\\
This means that we approximate Eq.~\eqref{schrf} to:
\begin{equation}\label{schrf2}
i\hbar\partial_t\vert\phi(t)\rangle = (\hat{H}_0+\hat{H}_p)\vert\phi(t)\rangle
\end{equation}
If the metric is random, Eq.~\eqref{schrf2} becomes a stochastic differential equation. As a consequence the predictions are given by taking the stochastic average over the random gravitational field. We then need to specify its stochastic properties.\\
As done for the bosonic particle case, we assume the noise to be gaussian and with zero mean. For the sake of simplicity, we also assume the different components of the metric fluctuation to be uncorrelated. This means that the noise is fully characterized by:
\begin{equation}\label{stocpropf}
\begin{split}
\mathbb{E} [ h_{\mu\nu}(\mathbf{x},t)] =& 0\\
\mathbb{E} [h_{\mu\nu}(\mathbf{x},t)h_{\mu\nu}(\mathbf{y},s)] =& \alpha^2f_{\mu\nu}(\mathbf{x},\mathbf{y};t,s)
\end{split}
\end{equation}
where we recall that $\mathbb{E}[\, \cdot \,]$ denotes the stochastic average, $\alpha$ represents the strength of the gravitational fluctuations, and $f(\mathbf{x},\mathbf{y};t,s)$ is the two point correlation function.\\
We move to the density operator formalism, and write the von Neumann equation for the averaged density matrix :
\begin{equation}
\begin{split}\label{vonf}
\partial_{t}\hat{\rho}(t)=& -\frac{i}{\hbar}\Big[ \hat{H}_0(t),\hat{\rho}(t)\Big]-\frac{i}{\hbar}\mathbb{E}\Big[ [\hat{H}_p(t),\hat{\Omega}(t)]\Big]\\
 \equiv & \mathbb{E}\Big[ \mathfrak{L}[\hat{\Omega(t)}]\Big]
\end{split}
\end{equation}
We solve the above equation perturbatively exploiting the cumulant expansion~\cite{vank} (see Appendix C of \cite{bosons}). With the further help of the gaussianity, zero mean, uncorrelation of different components, we can rewrite Eq.~\eqref{vonf}  in Fourier space~\footnote{Recall that our choice for the Fourier transform is: $$f(\mathbf{x}) = \frac{1}{(\sqrt{2\pi}\hbar)^{3}}\int d^3q\:\tilde{f}(\mathbf{q}) e^{i\mathbf{q}\cdot\mathbf{x}/\hbar} $$} as:
\begin{widetext}
\begin{equation}\label{nonmarkovf}
\begin{split}
&\partial_t \hat{\rho} =  -\frac{i}{\hbar}[\hat{H}_0,\hat{\rho}(t)]+\\
&-\frac{\alpha^2}{\hbar^8}\int \frac{d^3q\:d^3q^\prime}{(2\pi)^{3}}\int_{0}^{t}dt_{1}\:\frac{\tilde{f}^{00}(\mathbf{q},\mathbf{q^\prime};t,t_1)}{4}\Big[ \Big\lbrace e^{i\mathbf{q}\cdot\hat{\mathbf{x}}/ \hbar},\Xi_{00}(\hat{\mathbf{x}},\hat{\mathbf{p}})\Big\rbrace,\Big[\Big\lbrace e^{i\mathbf{q^\prime}\cdot\hat{\mathbf{x}}_{t_1}/\hbar},\Xi_{00}(\hat{\mathbf{x}}_{t_{1}},\hat{\mathbf{p}})\Big\rbrace ,\hat{\rho}(t)\Big]\Big]  \\
&  -\frac{\alpha^2c^2}{\hbar^8}\int\frac{d^3q\:d^3q^\prime}{(2\pi)^{3}}\int_{0}^{t}dt_{1}\:\frac{\tilde{f}^{0i}(\mathbf{q},\mathbf{q^\prime};t,t_1)}{4}\Big[\Big\lbrace e^{i\mathbf{q}\cdot\hat{\mathbf{x}}/ \hbar},\hat{p}_i\Big\rbrace,\Big[\Big\lbrace e^{i\mathbf{q^\prime}\cdot\hat{\mathbf{x}}_{t_1}/\hbar},\hat{p}_i\Big\rbrace,\hat{\rho}(t)\Big]\Big]\\
& -\frac{\alpha^2}{\hbar^8}\int \frac{d^3q\:d^3q^\prime}{(2\pi)^{3}}\int_{0}^{t}dt_{1} \:\frac{\tilde{f}^{ij}(\mathbf{q},\mathbf{q^\prime};t,t_1)}{4}\Big[ \Big\lbrace e^{i\mathbf{q}\cdot\hat{\mathbf{x}}/ \hbar},\Xi_{ij}(\hat{\mathbf{x}},\hat{\mathbf{p}})\Big\rbrace,\Big[\Big\lbrace e^{i\mathbf{q^\prime}\cdot\hat{\mathbf{x}}_{t_1}/\hbar},\Xi_{ij}(\hat{\mathbf{x}}_{t_{1}},\hat{\mathbf{p}})\Big\rbrace ,\hat{\rho}(t)\Big]\Big]\\
&+O(t\alpha^3\tau_c^2)\end{split}
\end{equation}
\end{widetext}
where we have introduced: 
\begin{equation}
\begin{split}
\Xi_{00} (\hat{\mathbf{x}},\hat{\mathbf{p}}) &= \frac{(\hat{\mathbf{p}}-\frac{e}{c}\mathbf{A}(t,\hat{\mathbf{x}}))^2}{4m}+\frac{m c^2}{2}-\frac{\hbar e}{2m c}\mathbf{B}(t,\hat{\mathbf{x}})\cdot\mathbf{\sigma}\\
\Xi_{ij}(\hat{\mathbf{x}},\hat{\mathbf{p}}) &= \frac{(\hat{p}_i-\frac{e}{c}A_i(t,\hat{\mathbf{x}}))(\hat{p}_j-\frac{e}{c}A_j(t,\hat{\mathbf{x}}))}{4m}+\frac{m c^2}{2}\\
&\quad+\frac{\hbar e}{2m c}\epsilon_{kil}{F^k}_j(t,\hat{\mathbf{x}})\sigma^l    
\end{split}
\end{equation}
for the sake of compactness, and $\hat{x}_{t_1} = e^{i\hat{H}_0t_{1}}\hat{x}e^{-i\hat{H}_0t_{1}}$.\\
The above equation describes the dynamics of a pointlike spin 1/2 fermionic particle in presence of an external weak, stochastic gravitational field (with the further assumptions made in this section) and an external electromagnetic field.\\
\\
We specialize Eq.~\eqref{nonmarkovf} to the Markovian limit, i.e. we assume the noise to be delta correlated in time, with the further assumptions of isotropy and homogeneity of the noise, so that its correlation function reads:
\begin{equation}
f^{\mu\nu}(\mathbf{x},\mathbf{y};t,s) = \lambda u^{\mu\nu}(\mathbf{x}-\mathbf{y})\delta(t-s)
\end{equation} 
where the factor $\lambda$ is in principle a generic coefficient with the dimension of a time. Note that the white noise assumption makes physical sense only if the correlation time ($\tau_{c}$) of the gravitational fluctuations is much smaller than the free dynamics' characteristic time  ($\tau_{free}$), or in the case where the contribution to the dynamics due to the gravitational perturbation is not affected by the free evolution dynamics, i.e. the operators describing the perturbation commute with the free dynamics operator $\hat{H}_0$. In such cases, as a first approximation, we can take $\lambda$ to be:
\begin{equation}\label{eq:lambda}
\lambda = \min(\tau_c \: ,\: t)
\end{equation}
Note that this choice does not affect the generality of the analysis as we leave $u^{\mu\nu}(\mathbf{x}-\mathbf{y})$ unspecified.\\
In such a regime Eq.~\eqref{nonmarkovf} is exact and it is easy to show that it reduces to:
\begin{widetext}
\begin{equation}\label{markovf}
\begin{split}
\partial_t \hat{\rho} = & -\frac{i}{\hbar}[\hat{H},\hat{\rho}(t)]\\
& -\frac{\alpha^2 \lambda}{(2\pi)^{3/2}\hbar^5 }\int d^3q \:\tilde{u}^{00}(\mathbf{q})\:\Big[\Big\lbrace e^{i\mathbf{q}\cdot\hat{\mathbf{X}}/ \hbar},\Xi_{00}(\hat{\mathbf{x}},\hat{\mathbf{p}})\Big\rbrace ,\Big[\Big\lbrace e^{-i\mathbf{q}\cdot\hat{\mathbf{X}}/\hbar},\Xi_{00}(\hat{\mathbf{x}},\hat{\mathbf{p}})\Big\rbrace ,\hat{\rho}(t)\Big]\Big]\\
&  -\frac{\alpha^2\lambda c^2}{(2\pi)^{3/2}\hbar^5 }\int d^3q \:\tilde{u}^{0i}(\mathbf{q})\:\Big[\Big\lbrace e^{i\mathbf{q}\cdot\hat{\mathbf{X}}/ \hbar},\hat{P}_i\big\rbrace,\Big[\Big\lbrace e^{-i\mathbf{q}\cdot\hat{\mathbf{X}}/\hbar},\hat{P}_i\Big\rbrace ,\hat{\rho}(t)\Big]\Big]\\
& -\frac{\alpha^2\lambda}{(2\pi)^{3/2}\hbar^5 }\int d^3q \:\tilde{u}^{ij}(\mathbf{q})\:\Big[\Big\lbrace e^{i\mathbf{q}\cdot\hat{\mathbf{X}}/ \hbar},\Xi_{ij}(\hat{\mathbf{x}},\hat{\mathbf{p}})\Big\rbrace,\Big[\Big\lbrace e^{-i\mathbf{q}\cdot\hat{\mathbf{X}}/\hbar},\Xi_{ij}(\hat{\mathbf{x}},\hat{\mathbf{p}})\Big\rbrace,\hat{\rho}(t)\Big]\Big]
\end{split}
\end{equation}
\end{widetext}
Eq.~\eqref{markovf} describes decoherence in a complex combination of position momentum and energy bases, as it contains double commutators of functions of the position, momentum and free kinetic energy operators with the averaged density matrix.\\
\\
In what follows we will specialize Eq.~\eqref{markovf} to determine under which approximations it recovers decoherence in the position or momentum eigenbasis only.\\
As for the bosonic case \cite{bosons}, the conditions:
\begin{eqnarray}
\left\{
\begin{array}{ll}
h^{00}\gtrsim h^{0i}\\
h^{00}\gtrsim h^{ij} \\
\Delta E \ll Mc^2\:(1-u^{00}(\Delta \mathbf{x})) 
\end{array}
\right.
\end{eqnarray}
are sufficient for our master equation to describe decoherence in the position eigenbasis only, where in this case the energy coherence needs to be modified to take into account the presence of the electromagnetic field, as $E= \frac{(\mathbf{p}-\frac{e}{c}\mathbf{A})^2}{2m}$. \\
Under the above assumptions, Eq.~\eqref{markovf} reads:
\begin{widetext}
\begin{equation}
\begin{split}
&\partial_t \hat{\rho} =  -\frac{i}{\hbar}[\hat{H},\hat{\rho}(t)] -\frac{\alpha^2 \lambda}{(2\pi)^{3/2}\hbar^5 }\int d^3q \:\tilde{u}^{00}(\mathbf{q})\Big[ e^{i\mathbf{q}\cdot\hat{\mathbf{X}}/ \hbar}\Big(\frac{mc^2}{2}-\frac{\hbar e \mathbf{\sigma}}{2m c}\cdot\mathbf{B}(t,\hat{\mathbf{x}})\Big),\Big[\Big\lbrace e^{-i\mathbf{q}\cdot\hat{\mathbf{X}}/\hbar}\Big(\frac{mc^2}{2}-\frac{\hbar e \mathbf{\sigma}}{2m c}\cdot\mathbf{B}(t,\hat{\mathbf{x}})\Big) ,\hat{\rho}(t)\Big]\Big]\\
\end{split}
\end{equation}
\end{widetext}
Contrary to the bosonic case, the condition of low momentum transfer
\begin{equation}
e^{i\mathbf{q}\cdot\hat{\mathbf{X}}/\hbar}\sim \hat{\mathbb{1}}
\end{equation}
is necessary but not sufficient to recover deocherence in the momentum or energy eigenbasis starting from Eq.~\eqref{markovf}. One in fact needs the further condition:
\begin{eqnarray}
\left\{
\begin{array}{ll}
|\mathbf{p}|\gg |\frac{e}{c}\mathbf{A}|\\
\frac{p^2}{2m}\gg|\frac{\hbar e \mathbf{\sigma}}{2mc}\cdot\mathbf{B}| \\
\end{array}
\right.
\end{eqnarray}
In this regime, Eq.~\eqref{markovf} can be approximated as:
\begin{equation}
\begin{split}
\partial_t \hat{\rho} = & -\frac{i}{\hbar}[\hat{H},\hat{\rho}(t)]\\
-& \frac{\alpha^2 \lambda}{(2\pi)^{3/2}\hbar^5 }\int d^3q \:\tilde{u}^{00}(\mathbf{q})\:\Big[\frac{\hat{\mathbf{p}}^2}{2m},\Big[\frac{\hat{\mathbf{p}}^2}{2m},\hat{\rho}(t)\Big]\Big]\\
-&  \frac{\alpha^2\lambda c^2}{(2\pi)^{3/2}\hbar^5 }\int d^3q \:\tilde{u}^{0i}(\mathbf{q})\:\Big[\hat{p}_i,\Big[\hat{p}_i,\hat{\rho}(t)\Big]\Big]\\
-& \frac{\alpha^2\lambda}{(2\pi)^{3/2}\hbar^5 }\int d^3q \:\tilde{u}^{ij}(\mathbf{q})\:\Big[\frac{\hat{p}_i\hat{p}_j}{2m},\Big[\frac{\hat{p}_i\hat{p}_j}{2m},\hat{\rho}(t)\Big]\Big]
\end{split}
\end{equation}
which indeed describes decoherence in the momentum eigenbasis.\\
\\
\section{Conclusions}
In this paper we have extended the results of our previous paper \cite{bosons} to the ferminonic case. We have  derived a
model of decoherence for non relativistic spin 1/2  particles interacting with both a weak stochastic gravitational perturbation and an external electromagnetic field. The resulting Hamiltonian and master equation correctly reproduce the results of the bosonic model in \cite{bosons} up to very small corrections. Such corrections account for relativistic effects (different spin of the two kind of particles) and the different quantization scheme employed.\\
The dynamics predicts also in this case decoherence in the position, momentum and energy eigenbasis, though under  different limiting cases than those described in \cite{bosons}.
\section*{Acknowledgments}
LA  thanks S. Ansoldi, L. Curcuraci, J.L. Gaona Reyes, C.I. Jones and S. Liberati for the helpful and inspiring discussions. The authors acknowledge financial support form the EU Horizon 2020 research and innovation program under Grant Agreement No. 766900 [TEQ].
GG acknowledges support from support from the Leverhulme Trust (RPG-2016- 046), the Spanish Agencia Estatal de Investigación, project PID2019-107609GB-I00 and QuantERA grant C'MON-QSENS!, by Spanish MICINN PCI2019-111869-2. LA thanks the University of Trieste and INFN. 
\appendix
\section{Vierbein (or tetrad) formulation of gravity}
We illustrate the basic ingredients of the tetrad formalism of the General Relativity theory. For a more complete treatment we address the reader to \cite{vierbein,gasp}.\\
\\
The standard geometrical interpretation of the gravitational interaction is based on the notion of the Riemannian metric ($g$) and the Christoffel connection ($\Gamma$). The spacetime curvature, its dynamical evolution and the interaction with matter sources are described through differential equations involving $g$ and $\Gamma$.\\
It is possible though to equivalently describe the geometry of a Riemannian maniflod ($M$) using the notion of vierbein and local connection. Such a formalism is particularly convenient when one wants to formulate a theory of gravity as a gauge theory, and wants to accomodate the notion of particles as irreducible representations of the Poincar\'{e} group in curved spacetimes \cite{kibble,sciama,hehl}. \\
We know that locally the laws of special relativity are valid. This translates into the consideration that we can attach at each and every point $p$ of the Riemannian manifold $M$ a flat tangent manifold equipped with the flat Minkowski metric.\\
There is a natural choice for the basis of such a tangent space ($T_{p} M$), the coordinate (or differential) basis:
\begin{equation}
    \hat{e}_{(\mu)}=\partial_{(\mu)}
\end{equation}
given by the partial derivatives of the coordinates. It follows that a given 4-vector $A\in T_{p} M$ has components:
\begin{equation}
    A=A^\mu\hat{e}_{(\mu)}=A^\mu\partial_{(\mu)}
\end{equation}
The dual basis:
\begin{equation}
    \hat{e}^{(\mu)}=dx^{(\mu)}
\end{equation}
spans the cotangent space, and it is given by the differential of the coordinates. A dual vector $B\in T_{p} M$ then has components:
\begin{equation}
B= B_{\mu}\hat{e}^{(\mu)}=B_\mu dx^{(\mu)}   
\end{equation}
As $T_{p} M$ is a vector space, we are in principle free to choose any orthonormal basis to span it, as long as $T_{p} M$ preserves the appropriate signature of the manifold. We therefore introduce a set of basis vectors $\hat{e}_a$, which we choose as non coordinate unit vectors, and we denote this choice by using small Latin letters for indices of the non coordinate frame. Such a non coordinate basis is called a tetrad basis. The condition for preserving the signature of the metric therefore reads:
\begin{equation}\label{innerproduct}
    g(\hat{e}_a,\hat{e}_b)=\eta_{ab}=\textrm{diag}(+,---)
\end{equation}
With this choice, we can clearly find a fixed orthonormal basis that is independent of position. Then, form a local prospective, any vector can be expressed as a linear combination of the fixed tetrad basis vectors at the point in the following way:
\begin{align}
{\hat{e}_{\mu}}(x)&={e_\mu}^a(x)\hat{e}_{(a)}  \\
V^a &={e_\mu}^aV^\mu
\end{align}
The 4x4 invertible matrix ${e_\mu}^a(x)$ is called a vierbein field (or tetrad), and it is the transformation matrix that maps the tangent space $T_{x} M$ into Minkowski space preserving the inner product.\\
The inverse vierbein field (or tetrad) has components ${e^\mu}_a(x)$, and satisfies the orthonormality condition:
\begin{align}
{e^\mu}_a  {e_{\nu}}^a &= \delta^\mu_\nu \nonumber\\    
{e_\mu}^a{e^\mu}_b &= \delta^a_b      
\end{align}
which come from the preservation of the inner product.\\
The vierbein fields are mixed indices objects, in the sense that they carry one Minkowski spacetime index $(a)$, and one Riemannian index ($\mu$). Accordingly, they transform under coordinate and Lorentz transformations respectively as:
\begin{align}
{e_\mu}^a   &\overset{\textrm{coord}}{\longrightarrow}  {e_\mu^\prime}^a= \frac{\partial x^\nu}{\partial x^{\prime \mu}}{e_\nu}^a    \\
{e_\mu}^a(x) &\overset{\textrm{Lorentz}}{\longrightarrow} {e_\mu^\prime}^a(x)= {\Lambda^a}_b {e_\mu}^b 
\end{align}
We now consider the covariant derivative  $\nabla X$ of a vector ($X$) in the Minkowski frame. It will be given by the standard derivative ($\partial X$) plus a correction given by the affine connection of the Minkowski frame:
\begin{equation}\label{connectionmink}
    (\nabla_\mu X^a) dx^\mu\otimes\hat{e}_{(a)}  = (\partial_\mu X^a + {{\omega_\mu}^a}_b X^b)dx^\mu\otimes\hat{e}_{(a)}
\end{equation}
The expression for the covariant derivative in the coordinate basis instead reads:
\begin{equation}\label{connectioncoord}
\begin{split}
\nabla X=&(\nabla_\mu x^\nu) dx^\mu\otimes\partial_{\nu}\\    
=&(\partial_\mu X^\nu +\Gamma_{\mu\alpha}^\nu X^\alpha)dx^\mu\otimes\partial_\nu\\
=& (\partial_\mu X^\nu +\Gamma_{\mu\alpha}^\nu X^\alpha)dx^\mu\otimes {e_{\nu}}^a (x)\hat{e}_{(a)}\\
=& {e_{\nu}}^a (x)\Big(\partial_\mu ({e^{\nu}}_b(x) X^b) +\Gamma_{\mu\alpha}^\nu{e^{\alpha}}_b(x)X^b\Big)dx^\mu\otimes\hat{e}_{(a)}
\end{split}
\end{equation}
Upon comparing Eq.~\eqref{connectionmink} with Eq.~\eqref{connectioncoord}, we can express the Minkowski frame or local affine connection in terms of the tetrads and the usual affine connection as:
\begin{equation}
{{\omega_\mu}^a}_b X^b = {e_\nu}^a(x) \partial_\mu {e^\nu}_b (x) X^b + {e_\nu}^a(x) {e^\alpha}_b(x)\Gamma_{\mu\alpha}^\nu X^b     
\end{equation}
Note that the above relation implies the metric compatibility condition:
\begin{align}
\nabla_\mu {e^\nu}_b(x) &= 0 \nonumber\\
\nabla_\mu g_{\alpha\beta}&= 0
\end{align}
Observing that $\nabla_\mu X^a$ must transform under a Lorentz boost as $X^a$, it follows:
\begin{equation}\label{lorentz}
\begin{split}
\nabla_\mu({\Lambda^a}_b)=&0\\  
=& \partial_\mu({\Lambda^a}_b)+{{\omega_\mu}^a}_c {\Lambda^c}_b- {{\omega_\mu}^c}_b {\Lambda^a}_c
\end{split}
\end{equation}
Upon multiplying the last line of Eq.~\eqref{lorentz} by ${\Lambda^b}_d$ on the left, we obtain the following relation:
\begin{equation}
{{\omega_\mu}^a}_d = {\Lambda^b}_d{\Lambda^a}_c{{\omega_\mu}^c}_b - {\Lambda^b}_d\partial_\mu({\Lambda^a}_b)     
\end{equation}
which tells us that the affine connection transforms inhomogeneously under Lorentz transformations.\\
One can construct the usual geometric objects from ($e$,$\omega$), as it is typically done from ($g$,$\Gamma$), such as the Curvature Tensor:
\begin{equation}
{R^{ab}}_{\mu\nu}= \partial_\mu {{\omega_\nu}^a}_b -  \partial_\nu {{\omega_\mu}^a}_b + {{\omega_\mu}^a}_c{{\omega_\nu}^c}_b- {{\omega_\nu}^a}_c{{\omega_\mu}^c}_b
\end{equation}
and the Torsion:
\begin{equation}
\mathfrak{T}_{\mu\nu}^a = \partial_\mu {e_\nu}^a - \partial_\nu {e_\mu}^a + {{\omega_\mu}^a}_b {e_\nu}^b - {{\omega_\nu}^a}_b {e_\mu}^b 
\end{equation}
The field equations for the vierbein field can be derived from a variational principle in the same fashion it is typically done for the metric. In order to show it, Let us recall the inner product-signature preservation condition Eq.~\eqref{innerproduct}, which can be equivalently recast into:
\begin{equation}
    g_{\mu\nu}= {e_{\mu}}^a\eta_{ab}{e_\nu}^b
\end{equation}
It then follows that the variation of the metric can be expressed in terms of the variation of the vierbein filed as:
\begin{equation}
\delta g_{\mu} = e_{\nu a}\delta {e_\mu}^a  +e_{\mu a} \delta {e_\nu}^a  = -(g_{\mu\lambda}{e_\nu}^a+g_{\nu\lambda}{e_\mu}^a) \delta {e_\lambda}^a
\end{equation}
The variation of the Einstein-Hilbert action ($S = \frac{1}{8\pi G}\int d^4 x \sqrt{-g} R $ \cite{mtw}) then reads:
\begin{equation}
\begin{split}
\partial_g S =& \frac{1}{8\pi G}\int d^4 x\: \sqrt{-g} (R^{\mu\nu}-\frac{1}{2}g^{\mu\nu}R)\delta g_{\mu\nu}\\
=& \frac{1}{8\pi G}\int d^4 x\: e(R^{\mu\nu}-\frac{1}{2}g^{\mu\nu}R) \frac{\partial g_{\mu\nu}}{\partial {e^\lambda}_a}\delta {e^\lambda}_a\\
=& \frac{1}{8\pi G}\int d^4 x\: e \Big({R_\lambda}^\nu {e_\nu}^a-\frac{1}{2}R\delta^\nu_\lambda{e_\nu}^a + {R^\mu}_\lambda {e_\mu}^a \\
&\quad\quad\quad\quad\quad\quad-\frac{1}{2}R\delta^\mu_\lambda{e_\mu}^a\Big)\delta {e^\lambda}_a\\
=& \frac{1}{8\pi G}\int d^4 x\: e \Big({R^\mu}_\nu-\frac{1}{2}\delta_\lambda^\mu R\Big) {e_\mu}^a \delta {e^\lambda}_a
\end{split} 
\end{equation}
Recalling the expression for the Einstein tensor ($G^{\mu\nu}=R^{\mu\nu}-\frac{1}{2}g^{\mu\nu}R$), the above equation yields:
\begin{equation}
G^{\mu\nu} {e_{\mu}}^a =0    
\end{equation}
which must be interpreted as the Einstein's equations for the vierbein field $e$. Note that in order to switch back to the usual metric formulation it is sufficient to multiply the above equation by ${e^\lambda}_a$.
We have thus shown that the vierbein formulation of General relativity is equivalent to the standard metric one.
\section{EOM for the spin 1/2 fermionic field}
We present the explicit steps for the derivation of the effective action of the fermionic action coupled to a weak gravitational perturbation.\\
\\
Consider the action for the Dirac field in curved spacetime:
\begin{equation}
\begin{split}
S = \int d^4x \sqrt{-g}\mathcal{L}_D
\end{split}
\end{equation}
with the Lagrangian density:
\begin{equation}
\mathcal{L}_{D}  = \frac{i\hbar c}{2}[ \bar{\psi}\gamma^{\mu}{e^{\small{A}}}_\mu\mathcal{D}_{\small{A}}\psi -{e^{\small{A}}}_\mu\mathcal{D}_{\small{A}}\bar{\psi}\gamma^{\mu}\psi  ] - mc^2 \bar{\psi}\psi
\end{equation}
where ${e^A}_\mu(x)$ is the so called vierbein field, 
the field that maps the tangent space to the manifold M at point $x$ $T_x M$ (coordinate basis $\partial_{\small{A}}$) into Minkowski space (non coordinate basis $\mathbf{e}_{\mu}$), and 
\begin{equation}
\mathcal{D}_{\mu} \psi= \partial_{\mu}\psi +\frac{1}{8}[\gamma_a,\gamma_b]{\omega_\mu}^{ab} \psi + \frac{ie}{\hbar c}A_{\mu}\psi
\end{equation}
is the covariant derivative with respect to both the spin and the electromagnetic connections. The pair $({e_A}^\mu,{\omega_A}^{\mu\nu})$ allows for an equivalent geometrization of the gravitational interaction to the standard one given in terms of the metric and the affine connection $(g_{AB},{\Gamma^{A}}_{BC})$; the relation between the two frameworks is given by:
\begin{eqnarray}\label{veirbein}
\left\lbrace
\begin{array}{cc}
{e_{A}}^\mu\eta_{\mu\nu}{e_{B}}^{\nu} =  g_{AB}\\
{\omega_{A}}^{\mu\nu} = {e_B}^{\mu}\eta^{\nu\rho}\partial_A {e^{B}}_\rho
+ {e_{B}}^{\mu}\eta^{\nu\rho} {e^{C}}_\rho{\Gamma^{B}}_{AC}
\end{array}
\right.
\end{eqnarray}  
Note that~\eqref{veirbein} holds only for a torsion free, metric compatible connection. \\
We write the metric as the sum of a flat background $\eta_{\mu\nu} = \text{diag}(+---)$, and a perturbation $h_{\mu\nu}$:
\begin{equation}
g_{\mu\nu} = \eta_{\mu\nu}+h_{\mu\nu}
\end{equation}
We are interested in studying the dynamics of the Dirac field in presence of a weak gravitational perturbation. We therefore perform a Taylor expansion of the action around the flat background metric and truncate the series at the first perturbative order:
\begin{equation}\label{acappr}
S \approx \int d^4 x\: (\sqrt{-g}\mathcal{L})\Big\vert_{g=\eta} - h^{\mu\nu}\Big(\frac{\partial (\sqrt{-g}\mathcal{L})}{\partial g^{\mu\nu}}\Big)\Big\vert_{g = \eta} + O(h^{2})
\end{equation}
In order to work out the explicit expression for $\frac{\partial (\sqrt{-g}\mathcal{L})}{\partial g^{\mu\nu}}$, we look at the variation of the action with respect to the metric tensor:
\begin{equation}\label{ferac}
\begin{split}
 \delta_{g}S &= -\frac{1}{2}\int d^4 x\: \sqrt{-g} T^{AB}\delta g_{AB}\\  
 &= \int d^4 x\: \frac{\partial (\sqrt{-g}\mathcal{L})}{\partial g_{AB}}\delta g_{AB}
\end{split}
\end{equation}
Notice that the above expression can be equivalently rewritten for a torsion free, metric compatible connection as:
\begin{equation}
\begin{split}
\delta_g S =& \int d^4 x\:  \frac{\partial (\sqrt{-g}\mathcal{L})}{\partial {e^{C}}_\alpha}\delta {e^{C}}_\alpha + \frac{\partial (\sqrt{-g}\mathcal{L})}{\partial \omega_{A\mu\nu}}\delta \omega_{A\mu\nu}\\
=& \int d^4 x\:  \sqrt{-g}\frac{\partial \mathcal{L}}{\partial {e^{C}}_\alpha}\delta {e^{C}}_\alpha + \sqrt{-g} \frac{\partial \mathcal{L}}{\partial \omega_{A\mu\nu}}\delta \omega_{A\mu\nu}\\
&+ \mathcal{L}\Big( \frac{\partial \sqrt{-g}}{\partial {e^{C}}_\alpha}\delta {e^{C}}_\alpha  + \frac{\partial \sqrt{-g}}{\partial \omega_{A\mu\nu}}\delta \omega_{A\mu\nu} \Big)
\end{split}
\end{equation} 
By noticing that $\frac{\partial \sqrt{-g}}{\partial \omega_{A\mu\nu}}=0$, and defining $\frac{\partial \mathcal{L}}{\partial {e^{C}}_\alpha}=:{\mathcal{T}_C}^\alpha$ and $\frac{\partial \mathcal{L}}{\partial \omega_{A\mu\nu}} =: \mathcal{S}^{A\mu\nu}$, we rewrite the above equation as:
\begin{equation}\label{ferac2}
\begin{split}
\delta S &= \int d^4x \sqrt{-g}\Big[{\mathcal{T}_C}^{\alpha} \delta {e^C}_{\alpha} + \mathcal{S}^{A\mu\nu}\delta\omega_{A\mu\nu} + 2{e_C}^\alpha \mathcal{L}_D \delta{e^C}_\alpha \Big]\\
&= \int d^4 x \sqrt{-g}\Big[\Big({\mathcal{T}_C}^{\alpha} + 2{e_C}^\alpha \mathcal{L}_D-\mathcal{D}_A[{{\mathcal{S}^{A}}_C}^{\alpha} - {{\mathcal{S}^{A}}^{\alpha}}_C \\ 
&+ {\mathcal{S}_C}^{A\alpha}
+ {\mathcal{S}^{\alpha A}}_{C} -{\mathcal{S}_C }^{\alpha A}- {{\mathcal{S}^{\alpha}}_C }^{A} ] \Big)\delta {e^C}_{\alpha} \Big]\\
&= \int d^4 x \sqrt{-g} ({B_C}^\alpha + 2{e_C}^\alpha \mathcal{L}_D) \delta {e^C}_{\alpha}\\
&=: \int d^4 x \sqrt{-g} {\Theta_C}^\alpha\delta {e^C}_{\alpha}
\end{split}
\end{equation}
where ${B_C}^\alpha$ is the Belinfante stress energy tensor \cite{belinfante}. In the case of a fermionic field it reads \cite{supergrav}:
\begin{equation}
\begin{split}
{B_C}^\alpha &= \frac{i\hbar c}{4} [ \bar{\psi}\gamma^{\alpha}\mathcal{D}_{C}\psi -\mathcal{D}_{C}\bar{\psi}\gamma^{\alpha}\psi + \bar{\psi}\gamma_{C}\mathcal{D}^{\alpha}\psi -\mathcal{D}^{\alpha}\bar{\psi}\gamma_{C}\psi]\\
&= \frac{1}{2}({\mathcal{T}_C}^{\alpha}+{\mathcal{T}^{\alpha}}_{C})
\end{split}
\end{equation}
Comparing Eq.~\eqref{ferac} and Eq.~\eqref{ferac2}, we notice:
\begin{equation}
\begin{split}
{\Theta_C}^\alpha \delta {e^C}_{\alpha} &=- \frac{1}{2}T^{AB}\delta g_{AB}\\
&= \frac{1}{2}T^{AB}(g_{AC}{e_B}^{\alpha} + g_{BC}{e_A}^{\alpha}) \delta {e^C}_{\alpha}\\
&=  {T_C}^{\alpha}\delta {e^C}_{\alpha}
\end{split}
\end{equation}
Thus we can write Eq.~\eqref{acappr} as:
\begin{equation}
\begin{split}
S &\approx \int d^4x \Big[(\sqrt{-g}\mathcal{L}_{D})\Big|_{g = \eta} + \frac{\partial(\mathcal{L}_D\sqrt{-g})}{\partial g_{AB}}\Big|_{g = \eta} h_{AB} + O(h^2)\Big]\\
& =  \int d^4 x \Big[(\sqrt{-g}\mathcal{L}_{D})\Big|_{g =\eta} -\frac{1}{2}(\Theta^{A\alpha}{e^B}_\alpha)\Big|_{g=\eta}h_{AB}+ O(h^2)\Big]\\
&=\int d^4 x \:\Big(\frac{i\hbar c}{2}[\bar{\psi}\gamma^{\mu}\nabla_\mu\psi - \nabla_\mu(\bar{\psi})\gamma^{\mu}\psi](1+\frac{tr(h)}{2})\\
&\quad-(1+\frac{tr(h)}{2})mc^2\bar{\psi}\psi- \frac{i\hbar c}{4}h_{\mu\nu}[\bar{\psi}\gamma^{\mu}\nabla^{\nu}\psi - \nabla^\nu(\bar{\psi})\gamma^{\mu}\psi ]\Big)\\
&\quad+ O(h^2)\\
&=: \int d^4 x\: \mathcal{L}_{eff} + O(h^2)
\end{split}
\end{equation}
and recover Eq.~\eqref{effectiveaction} of the main text.
\section{Foldy Wouthuysen method - fermionic model}
Here we illustrate the Fouldy Wouthuysen method applied to Eq.~\eqref{eqmotf}.
Let us consider the transformations: 
\begin{equation}
\mathfrak{H} \rightarrow  \mathfrak{H}^\prime = U(\mathfrak{H}-i\hbar\partial_t)U^{-1}
\end{equation}
and specialize $U$ to Eq.~\eqref{transf}, i.e.
\begin{equation}
U = e^{- i \gamma^0 \mathcal{O} /(2mc^2)} =: e^{iS}
\end{equation}
With the help of the BCH identity:
\begin{equation}
\begin{split}
\mathfrak{H}^\prime = & e^{iS} (\mathfrak{H}-i\hbar\partial_t) e^{-iS} = \mathfrak{H} + i [S,\mathfrak{H}] +\frac{i^2}{2!}[ S[S,\mathfrak{H}]] +\\
&+ \frac{i^3}{3!} [S[ S[S,\mathfrak{H}]]] + ...\\
&+\hbar(-\dot{S}-\frac{i}{2}[S,\dot{S}]+\frac{1}{6}[S,[S,\dot{S}]]+...)
\end{split}
\end{equation}
Recalling that:
\begin{equation}
\mathfrak{H} = mc^2\gamma^{0} + \mathfrak{E} + \mathcal{O}
\end{equation}
and noticing that: 
\begin{eqnarray}
\lbrack\gamma^0,\mathfrak{E}\rbrack &=& 0 \\
\lbrace\gamma^0 , \mathcal{O} \rbrace &=& 0 \\
\lbrack\gamma^0 \mathcal{O},\gamma^0\rbrack &=& -2\mathcal{O}\\
\lbrack\gamma^0 \mathcal{O}, \mathfrak{E}\rbrack &=& \gamma^0 \lbrack \mathcal{O},\mathfrak{E}\rbrack\\
\lbrack\gamma^0 \mathcal{O} ,\mathcal{O} \rbrack &=& 2 \gamma^0 \mathcal{O}^2
\end{eqnarray}
we get:
\begin{equation}
\mathfrak{H}^\prime = mc^2\gamma^0 +\mathfrak{E}^{\prime} + \mathcal{O}^\prime 
\end{equation}
where:
\begin{equation}\label{diag}
\begin{split}
\mathfrak{E}^\prime =& \mathfrak{E} + \gamma^0(\frac{\mathcal{O}^2}{2mc^2}-\frac{\mathcal{O}^4}{8m^3c^6})-\frac{1}{8m^2c^4}[\mathcal{O},[\mathcal{O},\mathfrak{E}]+i\hbar\dot{\mathcal{O}}]+...
\end{split}
\end{equation}
\begin{equation}
\mathcal{O}^\prime = \frac{1}{2mc^2}\gamma^0[\mathcal{O},\mathfrak{E}]-\frac{\mathcal{O}^3}{3m^2c^4}+\frac{i}{2mc^2}\gamma^0\dot{\mathcal{O}}+...
\end{equation}
We note that $\mathcal{O}^\prime$ is of order $c^{-1}$, meaning that we need to perform a further transformation if we want non trivial diagonal EOM. 
The transformation that we perform is:
\begin{equation}
U^\prime = e^{-i\gamma^0\mathcal{O}^\prime/(2mc^2)} 
\end{equation}
after which the Hamiltonian reads:
\begin{equation}
\mathfrak{H}^{\prime\prime} = mc^2\gamma^0 + \mathfrak{E}^\prime + \mathcal{O}^{\prime\prime}+...
\end{equation}
with:
\begin{equation}
\mathcal{O}^{\prime\prime}=\frac{\gamma^0}{2mc^2}[\mathcal{O}^\prime,\mathfrak{E}^\prime]+\frac{i}{2mc^2}\gamma^0\dot{\mathcal{O}^\prime}+...
\end{equation}
As $\mathcal{O}^{\prime\prime}\sim O(\frac{v^3}{c^3})$ we need to perform a final transformation:
\begin{equation}
U^{\prime\prime} = e^{-i\gamma^0\mathcal{O}^{\prime\prime}/(2mc^2)} 
\end{equation}
Finally the Hamiltonian reads:
\begin{equation}\label{appham}
H:= \mathfrak{H}^{\prime\prime\prime} = mc^2\gamma^0 + \mathfrak{E}^\prime +O(\frac{v^5}{c^5})
\end{equation}
In order to calculate the explicit expression of the Hamiltonian in Eq.~\eqref{appham}, we pick the Pauli representation for the Dirac gamma matrices:
\begin{equation}
    \gamma^0 = \left( \begin{array}{cc}
 \mathbb{1}  &\  \mathbb{0}    \\
  \mathbb{0} & -\mathbb{1} \end{array} \right),
\quad  \gamma^i = \left( \begin{array}{cc}
\mathbb{0} & \sigma^i    \\
-\sigma^i & \mathbb{0}\end{array} \right)\\
\end{equation}
\begin{equation}
    \alpha^i \equiv  \gamma^0\gamma^i =  \left( \begin{array}{cc}
\mathbb{0} & \sigma^i    \\
\sigma^i & \mathbb{0}\end{array} \right), \quad  \Sigma_i =  \left( \begin{array}{cc}
 \sigma_i  &\  \mathbb{0}    \\
  \mathbb{0} & \sigma_i \end{array} \right)
\end{equation}
By exploiting the identities:
\begin{align}\label{identities}
\Big[(\partial^j-\frac{ie}{\hbar c}A^j) ,(\partial_k-\frac{ie}{\hbar c}A_k) \Big] &= -\frac{ie}{\hbar c}{F^{j}}_{k}\\  
\alpha^i\alpha^j &= -\eta^{ij} + \epsilon^{ijk}\Sigma_k\nonumber\\
\lbrace\alpha^i,\alpha^j\rbrace &= -2\eta^{ij}\nonumber\\
\eta^{ij} &= -\delta^{ij}\nonumber
\end{align}
it only takes a bit of algebra to show that:  
\begin{widetext}
\begin{equation}
\begin{split}
\frac{\gamma^{0}}{2mc^2}\mathcal{O}^2 =& \frac{\gamma^{0}}{2mc^2} \Big(  - i\hbar c (1+\frac{h_{00}}{2}) (\partial_j - \frac{ie}{\hbar c}A_j)\gamma^0\gamma^j + \frac{i\hbar c}{2}h_{ij}(\partial^j - \frac{ie}{\hbar c}A^j)\gamma^0\gamma^i -\frac{i\hbar c}{4}\partial_i(\frac{tr(h)}{2}-h_{00})\gamma^0\gamma^i \\
&\quad+\frac{i\hbar}{4}\partial_t(h_{0i})\gamma^0\gamma^i  \Big)^2 \\
=& \gamma^{0}\Bigg[-\frac{\hbar^2}{2m}(1+h_{00})(\mathbf{\nabla}-\frac{ie}{\hbar c}\mathbf{A})^2 -\frac{\hbar e}{2m c}(1+h_{00})B^k\Sigma_k-\frac{\hbar^2}{2m}h_{ij}(\partial^i-\frac{ie}{\hbar c}A^i)(\partial^j-\frac{ie}{\hbar c}A^j)\\
&+\frac{\hbar e}{4 mc}\epsilon^{ijl}h_{jk}{F_i}^k\Sigma_l \Bigg] +\gamma^0 \Bigg[ \frac{\hbar^2}{2m}\partial^i(h_{00})\nabla_i -\frac{\hbar^2}{4m}\partial^i(h_{ij})\nabla^j -\frac{\hbar^2}{2m}\partial_i(\frac{tr(h)}{2}-h_{00})\nabla^i\\
&\quad\quad-\frac{i\hbar^2}{4m}\epsilon^{ijk}\Big(\partial_i(h_{00})\nabla_j-\partial_i(h_{jl})\nabla^l \Big)\Sigma_k -\frac{\hbar^2}{4m}\partial^i\partial_i (\frac{tr(h)}{2}-h_{00})\Bigg]+O(h^2)
\end{split}
\end{equation}
\end{widetext}
As the above term is of order $\gamma^{0}\frac{\mathcal{O}^2}{2mc^2}\sim O(\frac{v^2}{c^2})$, it follows that the next term in Eq.~\eqref{diag} is of order $\gamma^0\frac{\mathcal{O}^4}{8m^3c^6}\sim O(\frac{v^4}{c^4})$. After some algebra it reads:
\begin{widetext}
\begin{equation}
\begin{split}
&\mathcal{O}^4 = \Bigg(\hbar^2 c^2 (1+h_{00})(\mathbf{\nabla}-\frac{ie}{\hbar c}\mathbf{A})^2 -\hbar e c (1+h_{00})B^k\Sigma_k - \frac{i\hbar^2 c^2}{2}\epsilon^{ijk}\partial_i(h_{00})(\partial_j-\frac{ie}{\hbar c}A_j)\Sigma_k +\frac{\hbar^2 c^2}{2}\partial^i(h_{00})(\partial_i-\frac{ie}{\hbar c}A_i) \\
&- \hbar^2 c^2 h_{ij}(\partial^i-\frac{ie}{\hbar c}A^i)(\partial^j-\frac{ie}{\hbar c}A^j)+ \frac{\hbar e c}{2}\epsilon^{ijl}h_{jl}{F_i}^k\Sigma_l- \frac{\hbar^2}{4m}\partial^i(h_{ij})(\partial^j-\frac{ie}{\hbar c}A^j) + \frac{i\hbar^2 c^2}{2}\epsilon^{ijl}\partial_i(h_{jk})(\partial^k-\frac{ie}{\hbar c}A^k)\Sigma_l \\
&-\hbar^2 c^2 \partial_i(\frac{tr(h)}{2}-h_{00})\partial^i - \frac{\hbar^2 c^2}{2}\partial^i\partial_i (\frac{tr(h)}{2}-h_{00})\Bigg)^2   \\
&\quad\:= \hbar^4c^4(1+2h_{00})(\mathbf{\nabla}-\frac{ie}{\hbar c}\mathbf{A})^4 + \hbar^2 e c^2 (1+2h_{00})B^2 -\hbar^3 e c^3 (1+2h_{00})\lbrace(\mathbf{\nabla}-\frac{ie}{\hbar c}\mathbf{A})^2,B^k \rbrace\Sigma_k\\
&+\frac{\hbar^3 e c^3}{2}\epsilon^{ijl}\lbrace (\mathbf{\nabla}-\frac{ie}{\hbar c}\mathbf{A})^2 , h_{jk}{F_i}^k \rbrace \Sigma_l+ 2 \hbar^4 c^4 h_{ij}(\mathbf{\nabla}-\frac{ie}{\hbar c}\mathbf{A})^2(\partial^i-\frac{ie}{\hbar c }A^i)(\partial^j-\frac{ie}{\hbar c }A^j) - \frac{\hbar^3 e c^3}{2}\epsilon^{ijl}h_{jm}{F_i}^mB^k\lbrace \Sigma_k ,\Sigma_l\rbrace \\
&+ \frac{i\hbar^4 c^4}{2}\epsilon^{ijk}\lbrace (\mathbf{\nabla}-\frac{ie}{\hbar c}\mathbf{A})^2,\partial_i(h_{00})(\partial_j-\frac{ie}{\hbar c}A_j)\rbrace\Sigma_k - \frac{i\hbar^4 c^4}{2}\lbrace (\mathbf{\nabla}-\frac{ie}{\hbar c}\mathbf{A})^2,\partial^i(h_{00})(\partial_i-\frac{ie}{\hbar c}A_i)\rbrace  \\
&+\frac{\hbar^4 c^4}{2}\lbrace (\mathbf{\nabla}-\frac{ie}{\hbar c}\mathbf{A})^2,\partial_i(h_{ij})(\partial^j-\frac{ie}{\hbar c}A^j)\rbrace -\frac{\hbar^4 c^4}{2}\epsilon^{ijl}\lbrace (\mathbf{\nabla}-\frac{ie}{\hbar c}\mathbf{A})^2,\partial_i(h_{jk})(\partial^k-\frac{ie}{\hbar c}A^k)\rbrace\Sigma_l \\
&+\hbar^4 c^4 \lbrace (\mathbf{\nabla}-\frac{ie}{\hbar c}\mathbf{A})^2,\partial_i(\frac{tr(h)}{2}-h_{00})\partial_i\rbrace +\frac{\hbar^4 c^4}{2} \lbrace (\mathbf{\nabla}-\frac{ie}{\hbar c}\mathbf{A})^2,\partial^i\partial_i(\frac{tr(h)}{2}-h_{00})\rbrace + \frac{\hbar^3 e c^3}{2}\lbrace B^k,\partial^i(h_{ij})(\partial^j-\frac{ie}{\hbar c}A^j) \rbrace\Sigma_k \\
& + \frac{i\hbar^3 e c^3}{2}\epsilon^{ijl}\lbrace B^k\Sigma_k,\partial_i(h_{00})(\partial_j-\frac{ie}{\hbar c}A_j)\Sigma_l \rbrace - \frac{i\hbar^3 e c^3}{e}\lbrace B^k,\partial^i(h_{00})(\partial_i-\frac{ie}{\hbar c}A_i) \rbrace\Sigma_k\\
&+ \frac{i\hbar^3 e c^3}{2}\epsilon^{ijl}\lbrace B^k\Sigma_k,\partial_i(h_{jm})(\partial^m-\frac{ie}{\hbar c}A^m)\Sigma_l \rbrace +\hbar^3 e c^3\lbrace B^k,\partial_i(\frac{tr(h)}{2}-h_{00})\partial^i \rbrace \Sigma_k + \frac{\hbar^3 e c^3}{2}\lbrace B^k,\partial_i\partial^i(\frac{tr(h)}{2}-h_{00}) \rbrace \Sigma_k 
\end{split}    
\end{equation}
\end{widetext}
The last term in Eq.~\eqref{diag} requires lengthy intermediate calculations in order to get to the final result. We start by considering the expressions $[\mathcal{O},\mathfrak{E}]$ and $\dot{\mathcal{O}}$ separately. With the help of Eqs. \eqref{identities} and some algebra:
\begin{widetext}
\begin{equation}\label{EO}
\begin{split}
\Big[ \mathcal{O},\mathfrak{E} \Big] =& \Big[ - i\hbar c (1+\frac{h_{00}}{2}) (\partial_j - \frac{ie}{\hbar c}A_j)\gamma^0\gamma^j + \frac{i\hbar c}{2}h_{ij}(\partial^j -\frac{ie}{\hbar c}A^j)\gamma^0\gamma^i+\frac{i\hbar c}{4}\partial_t(h_{0i})\gamma^0\gamma^i+ \frac{i\hbar c}{4}\partial_i(\frac{tr(h)}{2}-h_{00})\gamma^0\gamma^i  \: \Large{\mathbf{,}}\\
& e A_0 +\frac{mc^2}{2}h_{00}\gamma^0 + i\hbar c \: h_{0i} (\partial^i-\frac{ie}{\hbar c}A^i)+\frac{i\hbar c}{4}\partial_{i}(h_0^i)+\frac{\hbar c}{4}\epsilon^{ijk}\partial_i(h_{0j})\Sigma_k -\frac{3i\hbar}{8}\partial_t(tr(h))+\frac{i\hbar}{4}\partial_t(h_{00}) \Big]\\
=& \:  i\hbar mc^3 h_{00}\nabla_i\gamma^i +\frac{i\hbar mc^3}{2}\partial_{i}(h_{00})\gamma^i-i\hbar e c\: h_{0j}{F_i}^j\alpha^i+\hbar^2c^2\partial_i(h_{0j})\nabla^j\alpha^i -i\hbar e c (1+\frac{h_{00}}{2}) \partial_i(A_0)\alpha^i\\ &+\frac{i\hbar e c}{2}h_{ij}\partial^j(A_0)\alpha^i  + \frac{\hbar^2 c^2}{2}\Big( \partial_j(h_{0i})\nabla^i\alpha^j-\partial_j(h_{0i}\nabla^j\alpha^i)\Big) - \frac{i\hbar^2 c^2}{4}\epsilon^{jkl}\partial_i\partial_j(h_{0k})\alpha^i\Sigma_l \\
&- \frac{3\hbar^2 c}{8}\partial_i\partial_t(tr(h))\alpha^i +\frac{\hbar^2 c}{4}\partial_i\partial_t(h_{00})\alpha^i 
\end{split}
\end{equation}
\end{widetext}
\begin{equation}\label{Odot}
\begin{split}
\dot{\mathcal{O}}=& -\frac{i\hbar c}{2}\partial_t(h_{00})\nabla_i\alpha^i \\
&- \frac{e}{2} h_{00}\partial_t(A_i)\alpha^i +\frac{i\hbar c}{2}\partial_t(h_{ij})\nabla^j\alpha^i +\frac{e}{2}h_{ij}\partial_t(A^j)\alpha^i\\
&\: + \frac{i\hbar c}{4}\partial_t\partial_i(\frac{tr(h)}{2}-h_{00})\alpha^i
\end{split}
\end{equation}
Upon plugging the Eq. (\ref{EO},\ref{Odot}) into the last term in Eq: \eqref{diag}, exploiting again the identities in Eqs. \eqref{identities}, and with a lot of algebra, we arrive at the final expression:
\begin{widetext}
\begin{equation}
\begin{split}
&-\frac{[\mathcal{O},[\mathcal{O},\mathfrak{E}]+i\hbar \dot{\mathcal{O}}]}{8m^2c^4}= +\frac{\hbar^2}{4m}h_{00}(\mathbf{\nabla}-\frac{ie}{\hbar c}\mathbf{A})^2\gamma^0 +\frac{e\hbar}{4mc}h_{00}\mathbf{B}\cdot\gamma^0\mathbf{\Sigma}+\frac{i\hbar^2 e}{4m^2c^2}(1+\frac{h_{00}}{2})\Big( \frac{\mathbf{\nabla}}{2}\times\mathbf{E} - \mathbf{E}\times\mathbf{\nabla}\Big)\cdot\boldsymbol{\Sigma}\\
&- (1+h_{00})\frac{\hbar^2e}{8m^2c^2}\mathbf{\nabla}\cdot\mathbf{E}-\frac{i\hbar^2 e}{16m^2c^2}\epsilon^{ikl}h_{ij}\partial^{j}(E_k)\Sigma_l -\frac{i\hbar^2e}{8m^2c^2}\epsilon^{ikl}h_{ij}E_k(\partial^j-\frac{ie}{\hbar c}A^j)\Sigma_l-\frac{\hbar^2 e}{8m^2c^2}h_{0j}\partial_i(F^{ij}) \\
&+ \frac{i\hbar^2 e}{4m^2c^2}\epsilon^{ijl}h_{0k}{F_j}^k(\partial_i-\frac{ie}{\hbar c}A_i)\Sigma_l+\frac{i\hbar^2 e}{8m^2c^2}\epsilon^{ijl}h_{0k}\partial_i({F_j}^k)\Sigma_l-\frac{\hbar^2 e}{16m^2c^2}\partial_i(h_{00})E^i -\frac{i\hbar^2 e}{16m^2c^2}\epsilon^{ijk}\partial_i(h_{00})E_j\Sigma_k\\
&-\frac{\hbar^2 e}{16m^2c^2}\partial^i(h_{ij})E^j +\frac{i\hbar^2 e}{16m^2c^2}\epsilon^{ijl}\partial_i(h_{jk})E^k\Sigma_l+\frac{\hbar^2}{8m}\partial_i(h_{00})(\partial^i-\frac{ie}{\hbar c}A^i)-\frac{i\hbar^2}{8m}\epsilon^{ijk}\partial_i(h_{00})(\partial_j-\frac{ie}{\hbar c}A_j)\Sigma_k\\
&-\frac{\hbar^2 e}{8m^2c^2}\partial^i(h_{0j}){F_i}^j+\frac{i\hbar^2 e}{8m^2c^2}\epsilon^{ijl}\partial_i(h_{0k}){F_j}^k\Sigma_l-\frac{i\hbar^2 e}{16m^2c^2}\epsilon^{ijk}\partial_i\left(\frac{tr(h)}{2}-h_{00}\right)E_j\Sigma_k\\
&-\frac{i\hbar^3}{16m^2c}\epsilon^{jkl}\epsilon^{ima}\partial_m(h_{0j})\lbrace (\partial_i-\frac{ie}{\hbar c}A_i), (\partial_k-\frac{ie}{\hbar c}A_k) \rbrace\Sigma_a\Sigma_l-\frac{\hbar^3}{8m^2c}\epsilon^{jkl}\partial^i(h_{0j})(\partial_k-\frac{ie}{\hbar c}A_k)(\partial_i-\frac{ie}{\hbar c}A_i)\Sigma_l\\ &-\frac{i\hbar^3}{8m^2 c}\epsilon^{jkl}{\epsilon_{l}}^{im}\partial_m(h_{0j})(\partial_k-\frac{ie}{\hbar c}A_k)(\partial_k-\frac{ie}{\hbar c}A_i)\Sigma_a\Sigma^a +\frac{i\hbar^2 e}{8m^2c^2}\epsilon^{jkl}\partial^i(h_{0j})F_{ki}\Sigma_l\\
&+\frac{\hbar^3}{16m^2c}\epsilon^{jkl}\partial^i\partial_i(h_{0j})(\partial_k-\frac{ie}{\hbar c}A_k)\Sigma_l-\frac{i\hbar^2 e}{16m^2c^2}\epsilon^{ikl}\partial_i(h_{0j}){F^j}_k\Sigma_l\\
&-\frac{i\hbar^3}{8m^2c}\partial^i\partial_i(h_{0j})(\partial^j-\frac{ie}{\hbar c}A^j)+\frac{\hbar^3}{8m^2c}\epsilon^{jki}\partial_i(h_{0j})(\partial_k-\frac{ie}{\hbar c}A_k)(\partial^l-\frac{ie}{\hbar c}A^l)\Sigma_l+\frac{i\hbar^3}{32m^2c^2}\epsilon^{ijl}\partial_i\partial_t(h_{jk})(\partial^k-\frac{ie}{\hbar c}A^k)\Sigma_l\\
&-\frac{\hbar^3}{16m^2 c}\epsilon^{kjl}\partial_k\partial_i(h_{0j})(\partial^i-\frac{ie}{\hbar c}A^i)\Sigma_l +\frac{h^3}{16m^2c}\epsilon^{jkl}\partial^i\partial_j(h_{0k})(\partial_i-\frac{ie}{\hbar c}A_i)\Sigma_l\\
&+\frac{i\hbar^3}{16m^2c}\epsilon^{jkl}\epsilon^{ima}\partial_m\partial_j(h_{0k})(\partial_i-\frac{ie}{\hbar c}A_i)\Sigma_l\Sigma_a - \frac{\hbar^3}{16m^2c}\epsilon^{jkl}\partial^i\partial_i\partial_j(h_{0k})\Sigma_l+\frac{i\hbar^3}{32m^2c^2}\partial^i\partial_i\partial_t(tr(h)-h_{00})\\
&-\frac{\hbar^2}{8m}\partial_i(h_{00})(\partial^i-\frac{ie}{\hbar c}A^i)\gamma^0 -\frac{\hbar^2}{16m}\partial^i\partial_i(h_{00})\gamma^0-\frac{i\hbar^3}{16m^2 c^2}\partial^i\partial_t(h_{00})(\partial^i-\frac{ie}{\hbar c}A^i)\\
&-\frac{\hbar^3}{16m^2c^2}\epsilon^{ijk}\partial_j\partial_t(h_{00})(\partial_i-\frac{ie}{\hbar c}A_i)\Sigma_k  -\frac{i\hbar^2 e}{8m^2c^3}\epsilon^{ijl}\partial_t(h_{jk}){F_i}^k\Sigma_l+\frac{i\hbar^3}{32m^2c^2}\partial^i\partial_t(h_{ij})(\partial^j-\frac{ie}{\hbar c}A^j)\\
&+\frac{i\hbar^3}{16m^2c^2}\epsilon^{ijk}\partial_t\partial_j(tr(h)-h_{00})(\partial_i-\frac{ie}{\hbar c}A_i)\Sigma_k +\frac{i\hbar^3}{32m^2c^2}\partial_t\partial^i(tr(h)-h_{00})(\partial_i-\frac{ie}{\hbar c}A_i)
\end{split}
\end{equation}
\end{widetext}
So that the total Hamiltonian reads:
\begin{widetext}
\begin{equation}
\begin{split}
    H=&  eA_0+\gamma^{0}\Bigg[mc^2(1+\frac{h_{00}}{2})-\frac{\hbar^2}{2m}(1+\frac{h_{00}}{2})(\mathbf{\nabla}-\frac{ie}{\hbar c}\mathbf{A})^2 -\frac{\hbar e}{2m c}(1+\frac{h_{00}}{2})B^k\Sigma_k-\frac{\hbar^2}{2m}h_{ij}(\partial^i-\frac{ie}{\hbar c}A^i)(\partial^j-\frac{ie}{\hbar c}A^j)\\
    &+\frac{\hbar e}{4 mc}\epsilon^{ijl}h_{jk}{F_i}^k\Sigma_l \Bigg]+\frac{i\hbar^2 e}{4m^2c^2}(1+\frac{h_{00}}{2})\Big( \frac{\mathbf{\nabla}}{2}\times\mathbf{E} - \mathbf{E}\times\mathbf{\nabla}\Big)\cdot\boldsymbol{\Sigma} - (1+h_{00})\frac{\hbar^2e}{8m^2c^2}\mathbf{\nabla}\cdot\mathbf{E}\\
&-\frac{i\hbar^2 e}{16m^2c^2}\epsilon^{ikl}h_{ij}\partial^{j}(E_k)\Sigma_l -\frac{i\hbar^2e}{8m^2c^2}\epsilon^{ikl}h_{ij}E_k(\partial^j-\frac{ie}{\hbar c}A^j)\Sigma_l\\ 
& + \frac{i\hbar^2 e}{4m^2c^2}\epsilon^{ijl}h_{0k}{F_j}^k(\partial_i-\frac{ie}{\hbar c}A_i)\Sigma_l-\frac{\hbar^2 e}{8m^2c^2}h_{0j}\partial_i(F^{ij})+\frac{i\hbar^2 e}{8m^2c^2}\epsilon^{ijl}h_{0k}\partial_i({F_j}^k)\Sigma_l\\
&-\frac{\gamma^0}{8m^3c^6}\Bigg[\hbar^4c^4(1+2h_{00})(\mathbf{\nabla}-\frac{ie}{\hbar c}\mathbf{A})^4 + \hbar^2 e c^2 (1+2h_{00})B^2 \\
&+ 2 \hbar^4 c^4 h_{ij}(\mathbf{\nabla}-\frac{ie}{\hbar c}\mathbf{A})^2(\partial^i-\frac{ie}{\hbar c }A^i)(\partial^j-\frac{ie}{\hbar c }A^j)+\frac{\hbar^3 e c^3}{2}\epsilon^{ijl}\lbrace (\mathbf{\nabla}-\frac{ie}{\hbar c}\mathbf{A})^2 , h_{jk}{F_i}^k \rbrace \Sigma_l \\
&- \frac{\hbar^3 e c^3}{2}\epsilon^{ijl}h_{jm}{F_i}^mB^k\lbrace \Sigma_k ,\Sigma_l\rbrace -\hbar^3 e c^3 (1+2h_{00})\lbrace(\mathbf{\nabla}-\frac{ie}{\hbar c}\mathbf{A})^2,B^k \rbrace\Sigma_k \Bigg]\\
&-\frac{\hbar^2}{8m}\partial_i(h_{00})(\partial^i-\frac{ie}{\hbar c}A^i)\gamma^0 -\frac{\hbar^2}{16m}\partial^i\partial_i(h_{00})\gamma^0 +\frac{i\hbar c}{4}\partial_{i}(h_0^i)\\
&+\frac{\hbar c}{4}\epsilon^{ijk}\partial_i(h_{0j})\Sigma_k -\frac{3i\hbar}{8}\partial_t(tr(h))+\frac{i\hbar}{4}\partial_t(h_{00})\\
& +\gamma^0 \Bigg[ \frac{\hbar^2}{2m}\partial^i(h_{00})\nabla_i -\frac{\hbar^2}{4m}\partial^i(h_{ij})\nabla^j -\frac{\hbar^2}{2m}\partial_i(\frac{tr(h)}{2}-h_{00})\nabla^i\\
&\quad\quad-\frac{i\hbar^2}{4m}\epsilon^{ijk}\Big(\partial_i(h_{00})\nabla_j-\partial_i(h_{jl})\nabla^l \Big)\Sigma_k -\frac{\hbar^2}{4m}\partial^i\partial_i (\frac{tr(h)}{2}-h_{00})\Bigg] \\
&+ H_{dd} +O(h^2) + O(\frac{v^5}{c^5})
\end{split}
\end{equation}
\end{widetext}
with
\begin{widetext}
{\allowdisplaybreaks
\begin{align}
& H_{dd} =   -\frac{\hbar^2 e}{16m^2c^2}\partial_i(h_{00})E^i -\frac{i\hbar^2 e}{16m^2c^2}\epsilon^{ijk}\partial_i(h_{00})E_j\Sigma_k-\frac{\hbar^2 e}{16m^2c^2}\partial^i(h_{ij})E^j +\frac{\hbar^2}{8m}\partial_i(h_{00})(\partial^i-\frac{ie}{\hbar c}A^i)\nonumber\\
&+\frac{i\hbar^2 e}{16m^2c^2}\epsilon^{ijl}\partial_i(h_{jk})E^k\Sigma_l-\frac{i\hbar^2}{8m}\epsilon^{ijk}\partial_i(h_{00})(\partial_j-\frac{ie}{\hbar c}A_j)\Sigma_k-\frac{\hbar^2 e}{8m^2c^2}\partial^i(h_{0j}){F_i}^j+\frac{i\hbar^2 e}{8m^2c^2}\epsilon^{ijl}\partial_i(h_{0k}){F_j}^k\Sigma_l\nonumber\\
&-\frac{i\hbar^2 e}{16m^2c^2}\epsilon^{ijk}\partial_i\left(\frac{tr(h)}{2}-h_{00}\right)E_j\Sigma_k-\frac{i\hbar^3}{16m^2c}\epsilon^{jkl}\epsilon^{ima}\partial_m(h_{0j})\lbrace (\partial_i-\frac{ie}{\hbar c}A_i), (\partial_k-\frac{ie}{\hbar c}A_k) \rbrace\Sigma_a\Sigma_l\nonumber\\ 
&-\frac{i\hbar^3}{8m^2 c}\epsilon^{jkl}{\epsilon_{l}}^{im}\partial_m(h_{0j})(\partial_k-\frac{ie}{\hbar c}A_k)(\partial_k-\frac{ie}{\hbar c}A_i)\Sigma_a\Sigma^a +\frac{i\hbar^2 e}{8m^2c^2}\epsilon^{jkl}\partial^i(h_{0j})F_{ki}\Sigma_l\nonumber\\
&-\frac{\hbar^3}{8m^2c}\epsilon^{jkl}\partial^i(h_{0j})(\partial_k-\frac{ie}{\hbar c}A_k)(\partial_i-\frac{ie}{\hbar c}A_i)\Sigma_l+\frac{\hbar^3}{16m^2c}\epsilon^{jkl}\partial^i\partial_i(h_{0j})(\partial_k-\frac{ie}{\hbar c}A_k)\Sigma_l-\frac{i\hbar^2 e}{16m^2c^2}\epsilon^{ikl}\partial_i(h_{0j}){F^j}_k\Sigma_l\nonumber\\
&-\frac{i\hbar^3}{8m^2c}\partial^i\partial_i(h_{0j})(\partial^j-\frac{ie}{\hbar c}A^j)+\frac{\hbar^3}{8m^2c}\epsilon^{jki}\partial_i(h_{0j})(\partial_k-\frac{ie}{\hbar c}A_k)(\partial^l-\frac{ie}{\hbar c}A^l)\Sigma_l\nonumber\\
&-\frac{\hbar^3}{16m^2 c}\epsilon^{kjl}\partial_k\partial_i(h_{0j})(\partial^i-\frac{ie}{\hbar c}A^i)\Sigma_l +\frac{h^3}{16m^2c}\epsilon^{jkl}\partial^i\partial_j(h_{0k})(\partial_i-\frac{ie}{\hbar c}A_i)\Sigma_l\nonumber\\
&+\frac{i\hbar^3}{16m^2c}\epsilon^{jkl}\epsilon^{ima}\partial_m\partial_j(h_{0k})(\partial_i-\frac{ie}{\hbar c}A_i)\Sigma_l\Sigma_a - \frac{\hbar^3}{16m^2c}\epsilon^{jkl}\partial^i\partial_i\partial_j(h_{0k})\Sigma_l\nonumber\\
&-\frac{\hbar^2}{8m}\partial_i(h_{00})(\partial^i-\frac{ie}{\hbar c}A^i)\gamma^0 -\frac{\hbar^2}{16m}\partial^i\partial_i(h_{00})\gamma^0-\frac{i\hbar^3}{16m^2 c^2}\partial^i\partial_t(h_{00})(\partial^i-\frac{ie}{\hbar c}A^i)\nonumber\\
&-\frac{\hbar^3}{16m^2c^2}\epsilon^{ijk}\partial_j\partial_t(h_{00})(\partial_i-\frac{ie}{\hbar c}A_i)\Sigma_k  -\frac{i\hbar^2 e}{8m^2c^3}\epsilon^{ijl}\partial_t(h_{jk}){F_i}^k\Sigma_l\nonumber\\
&+\frac{i\hbar^3}{16m^2c^2}\epsilon^{ijk}\partial_t\partial_j(tr(h)-h_{00})(\partial_i-\frac{ie}{\hbar c}A_i)\Sigma_k \nonumber\\
&+\frac{i\hbar^3}{32m^2c^2}\partial^i\partial_i\partial_t(tr(h)-h_{00})+\frac{i\hbar^3}{32m^2c^2}\partial^i\partial_t(h_{ij})(\partial^j-\frac{ie}{\hbar c}A^j)\nonumber\\
&+\frac{i\hbar^3}{32m^2c^2}\epsilon^{ijl}\partial_i\partial_t(h_{jk})(\partial^k-\frac{ie}{\hbar c}A^k)\Sigma_l+\frac{i\hbar^3}{32m^2c^2}\partial_t\partial^i(tr(h)-h_{00})(\partial_i-\frac{ie}{\hbar c}A_i)\nonumber\\
&-\frac{\gamma^0}{8m^3c^6} \Bigg[ \frac{i\hbar^4 c^4}{2}\epsilon^{ijk}\lbrace (\mathbf{\nabla}-\frac{ie}{\hbar c}\mathbf{A})^2,\partial_i(h_{00})(\partial_j-\frac{ie}{\hbar c}A_j)\rbrace\Sigma_k \nonumber\\
&- \frac{i\hbar^4 c^4}{2}\lbrace (\mathbf{\nabla}-\frac{ie}{\hbar c}\mathbf{A})^2,\partial^i(h_{00})(\partial_i-\frac{ie}{\hbar c}A_i)\rbrace \nonumber\\
&+ \frac{\hbar^4 c^4}{2}\lbrace (\mathbf{\nabla}-\frac{ie}{\hbar c}\mathbf{A})^2,\partial_i(h_{ij})(\partial^j-\frac{ie}{\hbar c}A^j)\rbrace \nonumber\\
&-\frac{\hbar^4 c^4}{2}\epsilon^{ijl}\lbrace (\mathbf{\nabla}-\frac{ie}{\hbar c}\mathbf{A})^2,\partial_i(h_{jk})(\partial^k-\frac{ie}{\hbar c}A^k)\rbrace\Sigma_l \nonumber\\
&+\hbar^4 c^4 \lbrace (\mathbf{\nabla}-\frac{ie}{\hbar c}\mathbf{A})^2,\partial_i(\frac{tr(h)}{2}-h_{00})\partial_i\rbrace\nonumber\\
& +\frac{\hbar^4 c^4}{2} \lbrace (\mathbf{\nabla}-\frac{ie}{\hbar c}\mathbf{A})^2,\partial^i\partial_i(\frac{tr(h)}{2}-h_{00})\rbrace + \frac{\hbar^3 e c^3}{2}\lbrace B^k,\partial^i(h_{ij})(\partial^j-\frac{ie}{\hbar c}A^j) \rbrace\Sigma_k \nonumber\\
& + i\hbar^3 e c^3\epsilon^{ijl}\lbrace B^k\Sigma_k,\partial_i(h_{00})(\partial_j-\frac{ie}{\hbar c}A_j)\Sigma_l \rbrace - i\hbar^3 e c^3\lbrace B^k,\partial^i(h_{00})(\partial_i-\frac{ie}{\hbar c}A_i) \rbrace\Sigma_k\nonumber\\
&+ i\hbar^3 e c^3\epsilon^{ijl}\lbrace B^k\Sigma_k,\partial_i(h_{ja})(\partial^a-\frac{ie}{\hbar c}A^{a})\Sigma_l \rbrace +\hbar^3 e c^3\lbrace B^k,\partial_i(\frac{tr(h)}{2}-h_{00})\partial^i \rbrace \Sigma_k \nonumber\\
&+ \frac{\hbar^3 e c^3}{2}\lbrace B^k,\partial_i\partial^i(\frac{tr(h)}{2}-h_{00}) \rbrace \Sigma_k\Bigg]
\end{align}
}
\end{widetext}
By neglecting the terms containing derivatives of the gravitational field of order $\frac{v^3}{c^3}$ or higher (namely the term $H_{dd}$), we recover Eq.~\eqref{hamf} of the main text.
     \bibliographystyle{unsrt}
\bibliography{biblio}

\begin{thebibliography}{10}

\bibitem{bosons}
L.~Asprea G. Gasbarri~A. Bassi.
\newblock Gravitational decoherence: a general non relativistic model.
\newblock {\em arXiv:1905.01121v2}, 2019.

\bibitem{ligo}
The~LIGO Collaboration and the VIRGO~Collaboration.
\newblock Observation of gravitational waves from a binary black hole merger.
\newblock {\em Physical Review Letters}, 116:061102, Feb 2016.

\bibitem{neutronstar}
The~LIGO Collaboration and the VIRGO~Collaboration.
\newblock Gw170817: Observation of gravitational waves from a binary neutron
  star inspiral.
\newblock {\em Physical Review Letters}, 119:161101, Oct 2017.

\bibitem{tian}
J.~Luo et~al.
\newblock {TianQin}: a space-borne gravitational wave detector.
\newblock {\em Classical and Quantum Gravity}, 33(3):035010, jan 2016.

\bibitem{decigo}
S.~Kawamura et~al.
\newblock The japanese space gravitational wave antenna - {DECIGO}.
\newblock {\em Journal of Physics: Conference Series}, 122:012006, jul 2008.

\bibitem{et}
M.~Punturo et~al.
\newblock The {Einstein} telescope: a third-generation gravitational wave
  observatory.
\newblock {\em Classical and Quantum Gravity}, 27(19):194002, sep 2010.

\bibitem{kagra}
The~KAGRA Collaboration.
\newblock {Construction of KAGRA: an underground gravitational-wave
  observatory}.
\newblock {\em Progress of Theoretical and Experimental Physics}, 2018(1), 01
  2018.

\bibitem{lisa}
K.~Danzmann and the LISA~study team.
\newblock {LISA}: laser interferometer space antenna for gravitational wave
  measurements.
\newblock {\em Classical and Quantum Gravity}, 13(11A):A247--A250, Nov 1996.

\bibitem{aha}
A.~Stern Y.~Imry, Y.~Aharonov.
\newblock Phase uncertainty and loss of interference: A general picture.
\newblock {\em Physical Review A}, 41:3436--3448, Apr 1990.

\bibitem{linet}
B.~Linet~P. Tourrenc.
\newblock Changement de phase dans un champ de gravitation- possibilité de
  détection interférentielle.
\newblock {\em Canadian Journal of Physics}, 54, 1976.

\bibitem{goku}
E.~G\"{o}kl\"{u} C.~L\"{a}mmerzahl.
\newblock Metric fluctuations and the weak equivalence principle.
\newblock {\em Classical and Quantum Gravity}, 25(10):105012, May 2008.

\bibitem{breuer}
H.P.~Breuer C.~L\"{a}mmerzahl, E.~G\"{o}kl\"{u}.
\newblock Metric fluctuations and decoherence.
\newblock {\em Classical and Quantum Gravity}, 26(10):105012, Apr 2009.

\bibitem{sanchez}
J.L.~Sanchez Gomez.
\newblock Decoherence through stochastic fluctuations of the gravitational
  field.
\newblock {\em (ed.L. Diosi, B. Lukacs), pp. 88-93. Singapore: World
  Scientific}, 456:88--93, 1992.

\bibitem{power}
W.L.~Power I.C.~Percival.
\newblock Decoherence of quantum wave packets due to interaction with conformal
  space-time fluctuations.
\newblock {\em Proceedings: Mathematical, Physical and Engineering Sciences},
  456:955--968, 2000.

\bibitem{blencowe}
M.P. Blencowe.
\newblock Effective field theory approach to gravitationally induced
  decoherence.
\newblock {\em Physical Review Letters}, 111:021302, Jul 2013.

\bibitem{ana}
B.~L.~Hu C.~Anastopoulos.
\newblock A master equation for gravitational decoherence: probing the textures
  of spacetime.
\newblock {\em Classical and Quantum Gravity}, 30(16):165007, Jul 2013.

\bibitem{lamine}
B.~Lamine S.~Reynaud, M.T.~Jaekel.
\newblock Gravitational decoherence of atomic interferometers.
\newblock {\em The European Physical Journal D - Atomic, Molecular, Optical and
  Plasma Physics}, 20(2):165--176, Aug 2002.

\bibitem{bd}
N.~D. Birrell and P.~C.~W. Davies.
\newblock {\em Quantum Fields in Curved Space}.
\newblock Cambridge Monographs on Mathematical Physics. Cambridge University
  Press, 1982.

\bibitem{vierbein}
J.~Yepez.
\newblock Einstein's vierbein field theory of curved space.
\newblock {\em arXiv:1106.2037v1 [gr-qc]}, 2011.

\bibitem{greiner}
W.~Greiner.
\newblock {\em Relativistic quantum mechanics.}
\newblock Springer-Verlag Berlin Heidelberg, 2000.

\bibitem{foldy}
L.L.~Foldy S.~A.Wouthuysen.
\newblock On the dirac theory of spin 1/2 particles and its non-relativistic
  limit.
\newblock {\em Physical Review}, 78:29--36, Apr 1950.

\bibitem{armin}
A.~Watcher.
\newblock {\em Relativistic quantum mechanics}.
\newblock Theoretical and mathematical physics - Springer, 2011.

\bibitem{mtw}
J.~Wheeler C.~Misner, K.~Thorne.
\newblock {\em Gravitation}.
\newblock Freeman, New York, 1973.

\bibitem{vank}
N.G.~Van Kampen.
\newblock {\em Stochastic processes in physics and chemistry}.
\newblock North Holland Personal Library, Elevier, 1981.

\bibitem{gasp}
V.~de~Sabba M.~Gasperini.
\newblock {\em Introduction to gravitation}.
\newblock World Scientific, 1985.

\bibitem{kibble}
T.W.B. Kibble.
\newblock Lorentz invariance and the gravitational field.
\newblock {\em Journal of Mathematical Physics}, 2(212), 1961.

\bibitem{sciama}
D.W. Sciama.
\newblock On the analogy between charge and spin in general relativity.
\newblock {\em in Recent Developments in General Relativity - Pergamon+ PWN,
  Oxford}, 1962.

\bibitem{hehl}
F.W. Hehl~P. von~der Heyde G.D.~Kerlick.
\newblock General relativity with spin and torsion: Foundations and prospects.
\newblock {\em Reviews of Modern Physics}, 48(3), Jul 1976.

\bibitem{belinfante}
F.J. Belinfante.
\newblock On the current and the density of the electric charge, the energy,
  the linear momentum and the angular momentum of arbitrary fields.
\newblock {\em Physica}, 7:449--474, May 1940.

\bibitem{supergrav}
D.Z. Freedman A.~Van Proeyen.
\newblock {\em Supergravity.}
\newblock Cambridge University Press, 2012.

\end{thebibliography}
\end{document}